\documentclass[article]{aa}
\usepackage{graphicx}
\usepackage{txfonts}
\usepackage{natbib}
\usepackage[colorlinks=true,citecolor=blue]{hyperref}
\usepackage{xspace}
\usepackage{subfig}
\usepackage{xcolor}

\usepackage[T1]{fontenc}

\bibpunct{(}{)}{;}{a}{}{,}

\newcommand{\MJup}{\ensuremath{M_{\mathrm{Jup}}}\xspace}
\newcommand{\RJup}{\ensuremath{R_{\mathrm{Jup}}}\xspace}

\newcommand{\as}{\hbox{$^{\prime\prime}$}\xspace}

\begin{document}

\title{The widest H$\alpha$ survey of accreting protoplanets around nearby transition disks \thanks{Based on observations collected at the European Southern Observatory (ESO), programs number 096.C-0267, 096.C-0248, 099.C-0453, 0100.C-0193, 0101.C-0461, 0102.C-0138.}}

   \author{A. Zurlo\inst{1,2}, G. Cugno\inst{3}, M. Montesinos\inst{4,5,6}, S. Perez\inst{7}, H. Canovas\inst{8}, S. Casassus\inst{9}, V. Christiaens\inst{10}, L. Cieza\inst{1},\newline  N. Huelamo\inst{11} }

   \institute{\inst{1}N\'ucleo de Astronom\'ia, Facultad de Ingenier\'ia y Ciencias, Universidad Diego Portales, Av. Ejercito 441, Santiago, Chile\\
     \inst{2}Escuela de Ingenier\'ia Industrial, Facultad de Ingenier\'ia y Ciencias, Universidad Diego Portales, Av. Ejercito 441, Santiago, Chile \\
     \inst{3}ETH Z\"urich, Institute for Particle Physics and Astrophysics, Wolfgang-Pauli-Str. 27, 8093 Zürich, Switzerland \\
\inst{4}N\'ucleo Milenio de Formaci\'on Planetaria (NPF), Chile \\
     \inst{5}Instituto de F\'isica y Astronom\'ia, Universidad de Valpara\'iso, Valpara\'iso, Chile \\
     \inst{6}Chinese Academy of Sciences South America Center for Astronomy, National Astronomical Observatories, CAS, Beijing, China \\
\inst{7}Universidad de Santiago de Chile, Av. Libertador Bernardo O'Higgins 3363, Estacion Central, Santiago, Chile \\
       \inst{8}European Space Astronomy Centre (ESA/ESAC), Operations Department, Villanueva de la Ca\~nada, Madrid, Spain \\
     \inst{9}Departamento de Astronom\'ia, Universidad de Chile, Casilla 36-D, Santiago, Chile\\
     \inst{10}School of Physics and Astronomy, Monash University, VIC 3800, Australia \\
     \inst{11}Centro de Astrobiolog\'{\i}a (CSIC-INTA), ESAC, Camino bajo del Castillo s/n, E-28692 Villanueva de la Ca\~nada, Madrid, Spain 
   }         
   \date{Submitted 2019 October, 10 / Accepted 2019 December, 9}


  \abstract
 {The mechanisms of planet formation are still under debate. We know little about how planets form, even if more than 4000 exoplanets have been detected to date. Recent investigations target the cot of newly born planets: the protoplanetary disk. At the first stages of their life, exoplanets still accrete material from the gas-rich disk in which they are embedded. Transitional disks are indeed disks that show peculiarities, such as gaps, spiral arms, and rings, which can be connected to the presence of substellar companions. }
    {To investigate what is responsible for these features, we selected all the known transitional disks in the solar neighborhood (<200 pc) that are visible from the southern hemisphere. We conducted a survey of 11 transitional disks (TDs) with the SPHERE instrument at the Very Large Telescope. This is the largest H$\alpha$ survey that has been conducted so far to look for protoplanets. The observations were performed with the H$\alpha$ filter of ZIMPOL in order to target protoplanets that are still in the accretion stage. All the selected targets are very young stars, less than 20 Myr, and show low extinction in the visible. }
    {We reduced the ZIMPOL pupil stabilized data by applying the method of the angular spectral differential imaging (ASDI), which combines both techniques. The datacubes are composed of the Cnt\_H$\alpha$ and the narrow band filter H$\alpha$, which are taken simultaneously to permit the suppression of the speckle pattern. The principal component analysis (PCA) method was employed for the reduction of the data. For each dataset, we derived the 5$\sigma$ contrast limit and converted it in upper limits on the accretion luminosity.   }
   {We do not detect any new accreting substellar companions around the targeted transition disks down to an average contrast of 12 magnitudes at 0\farcs2 from the central star. We have recovered the signal of the accreting M star companion around the star HD\,142527. We have detected and resolved, for the first time in visible light, the quadruple system HD\,98800. For every other system, we can exclude the presence of massive actively accreting companions, assuming that the accretion is not episodic and that the extinction is negligible. The mean accretion luminosity limit is 10$^{-6}$ L$_{\odot}$ at a separation of 0\farcs2 from the host.  }
   {}

   \keywords{Instrumentation: high angular resolution, high contrast, Methods: data analysis, Techniques: imaging, Stars: AKSco, V4046Sgr, HD169142, HD98800B, HD97048, HD100546, HD142527, HD141569A, HD135344B, PDS66, HD100453, UXTauA}

\titlerunning{H$\alpha$ survey of close TDs}
\authorrunning{Zurlo et al.}
\maketitle
%
\section{Introduction}

Detecting forming protoplanets is the cornerstone to understanding planet formation \citep{2008ApJ...678.1419F,2012ApJ...745..174S,2017A&A...608A..72M}. Approximately $\sim$30\% of Herbig Ae/Be disks should host giant planets of $\sim$0.1 to 10~M$_{Jup}$ \citep{2015A&A...582L..10K}. Some Herbig stars host transitional disks (TDs), which are peculiar disks with cavities, gaps, and spiral structures that can be induced by the presence of a companion \citep[e.g.,][]{2011ApJ...738..131D}.  Supporting this theory, hydrodynamical simulations predicted the presence of substellar companions around some observed TDs \citep[e.g.,][]{2016ApJ...826...75D}.

Accreting giant planets are thought to develop a circumplanetary disk (CPD) as they interact with the circumstellar material \citep[e.g.,][]{Miki1982, Gressel2013, Perez2015}. While a fraction of the mass flows through the CPD at midplane latitudes, a substantial amount of gas falls onto the surface of the CPD via meridional flows \citep{Tanigawa2012, Morbidelli2014, Szulagyi2014}. The vertical accretion flow happens at near free-fall speeds, thus shocking the CPD surface and heating the gas to thousands of kelvin.  H$\alpha$ emission is thought to arise from the hot, shocked surface \citep{Aoyama2018}. Additionally, the intrinsic planet luminosity also contributes to heat up the circumplanetary environment, while enhancing the stellocentric accretion rate \citep{2015ApJ...806..253M}.

\newpage

Breakthroughs in H$\alpha$ high-contrast imaging, which is an indicator of accretion on compact bodies \citep{2012A&A...548A..56R, 2014A&A...562A.104T, 2015Natur.527..310Z}, have recently triggered intense interest. Notably, the first successful
H${\alpha}$ detection of a close stellar companion was found around the star HD142527 \citep{2014ApJ...781L..30C}, and was followed by the detection of an accreting protoplanet candidate around the transition disk LkCa15 \citep{2015Natur.527..342S}. The emission from the potential protoplanet is at the same location of the scattered light of the interior part of the disk \citep{2016ApJ...828L..17T}, and it is hard to disentangle the two components. The presence of such a protoplanet is a subject of debate \citep[see][]{2018A&A...618L...9M, 2019ApJ...877L...3C}. The most evident case of an accreting protoplanet in a transition disk is PDS70b, where a clear detection in the near-infrared (NIR) has been recovered \citep{2018A&A...617A..44K,2018A&A...617L...2M, 2019ApJ...877L..33C, 2019A&A...632A..25M}. Subsequently, \cite{2018ApJ...863L...8W} found H$\alpha$ emission coming from the location of the protoplanet, demonstrating that the object is currently actively accreting with a mass accretion rate of $\dot{M}$ = 10$^{-8 \pm 1}$ \MJup yr$^{-1}$. Recently, a second accreting protoplanet around the same star was detected in H$\alpha$ emission with the multi-unit spectroscopic explorer (MUSE) instrument by \citet{2019NatAs...3..749H}. Finally, \citet{2019A&A...622A.156C} published a first H$\alpha$ survey focusing on targets that are suspected of hosting forming companions; they were unable to find any new unknown accretion signals.

Imaging young (a few Myr old) disks in H$\alpha$ could potentially be sensitive to low-mass planets, even if they have moderate accretion rates \citep[$\dot{M}$ $>$ 10$^{-5}$ \MJup yr$^{-1}$;][]{2014ApJ...781L..30C}. These values were estimated by converting the accretion rate of a low-mass object to H$\alpha$ luminosity by using the TTauri stars relation presented in \cite{2012A&A...548A..56R} in order to predict the contrast in H$\alpha$, showing that observations are more favorable for planetary objects with respect to any other NIR observation \citep[Fig. 5 of][]{2014ApJ...781L..30C}. This is the case even though the two recent papers by \cite{2019ApJ...885...94T} and \citet{2019ApJ...885L..29A} demonstrate that the H$\alpha$ emission from the planets comes from different mechanisms.

For this purpose, we selected all the known TDs that are closer than 200 pc and observable with the SPHERE planet finder at the Very Large Telescope (VLT, Chile). We found 11 targets, and for completeness we added three targets from the archive. The selected stars are young (age $<$ 20 Myr), and all of them have very low extinction in the visible (a fraction of a magnitude). Our targets are young stars with disks that present spirals, cavities, or gaps, and even in some cases, in both gas and dust. Material ought to flow from the outer disk reservoir and cross the gap to replenish the material being accreted by these stars. Hydro simulations suggest that the most obvious gap crossing mechanism is through the protoplanet's wake streams.

The outline of the paper is as follows. In Sec.~\ref{sec:obs}, we describe our target selection and present the SPHERE H$\alpha$ observations of the systems; in Sec.~\ref{sec:red}, we describe the reduction methods applied and the results that we obtained in Sec.~\ref{sec:anal}.  We give the discussion and conclusions in Sec.~\ref{sec:conc}.


\section{Observations}
\label{sec:obs}

\subsection{Target selection}
Our survey targets all the known TDs that are closer than 200 pc and observable from the VLT. Some of them had already been observed in a similar configuration of the instrument and for the same purpose. The data were already available in the archive. The individual target PDS 70, which was initially selected as part of the sample before the discovery of the companions b and c \citep{2018A&A...617A..44K,2019NatAs...3..749H}, has been removed from this analysis, as it will be presented in a separate and future paper. The stars included in the sample are listed - together with their main properties - in Table~\ref{t:sample}.

\begin{center}
  \begin{table*}
    \caption{List of objects included in the sample with their main characteristics. The objects marked with $^{*}$ have been included for completeness, but they are a part of other observing programs. All the distances are taken from \citet{2018yCat.1345....0G} apart, from the one of HD\,98800B, which is from \citet{2007A&A...474..653V}. References on the peculiarities of each disk are found in the main text. }
    \label{t:sample}
    \begin{minipage}{1\textwidth}
    \begin{tabular}{lcccccc}
\hline
 Name & Sp. Type & d (pc)  & Age (Myr) & Planet formation/presence indicators \\
\hline
MWC\,758$^{*}$ &   A8Ve       & 160  &  1.5\footnote{\citet{2012A&A...544A..78M}}   &  Spiral arms, possibly shaped by a $\sim 10$ M$_{Jup}$ companion at 160 au  \\
HD\,135344\,B$^{*}$ & F8V  & 136 &  8\footnote{\citet{2009ApJ...699.1822G}}  &     Possible planet induced spiral features     \\
HD\,142527$^{*}$ & F6III & 157  & 5\footnote{\citet{2014ApJ...790...21M}} &  Accreting stellar companion, inner warp, outer two-arms spirals disk   \\
 HD\,100546 & B9V  & 110   &  5\footnote{\citet{2015MNRAS.453..976F}}      &    Spirals, protoplanet candidates, bar-like structure    \\
HD\,98800\,B & K5 & 47 & 7--10\footnote{\citet{2014A&A...563A.121D}} & {Gap from 2 to 5.9 au } \\
AK\,Sco & Ae  & 141 & 10--20\footnote{\citet{2012ApJ...746..154P,2012AJ....144....8S}} & Two spiral arms with extension of 13-40 au  \\ 
V4046\,Sgr\,AB & K5V & 72  &10--20\footnote{\citet{2019NatAs...3..167D}} & Rings at 14, 29, and 40 au \\  
HD\,169142 & B5V & 114  & 3--12\footnote{\citet{2007ApJ...665.1391G}} & Gap at 23 and 40 au \\  
DoAr\,44 &  K3  & 146 & 7\footnote{\citet{2009ApJ...700.1502A}}  & Gap from 5 to 32 au, a rescaled version of HD\,142527, {inner warp }\\ 
HD\,97048 & A0V & 185  & 2--6\footnote{\citet{1998A&A...330..145V,2006Sci...314..621L,2019ApJ...872..112V}} & Gaps at 34, 79, 140 au, and 179 au and 4 rings \\
HD\,141569\,A & A0Ve & 106  & 5\footnote{\citet{2004A&A...419..301M}}  & Ringlets at 47 au, 64 au, and 93 au \\   
UX\,Tau\,A &  K2Ve    &  140  &   1\footnote{\citet{2008ApJS..176..184F}}    & Gap at 56 au          \\
HD\,100453 & A9Ve  &  104  & 10\footnote{\citet{2009ApJ...697..557C}} & Inner cavity up to 19 au  \\      
PDS\,66  &  K1Ve   & 86  &  17\footnote{\citet{2002AJ....124.1670M}}  & Inner cavity up to 15 au  \\         
\hline
    \end{tabular}
    \end{minipage}
\end{table*}
\end{center}

The disks around MWC\,758, HD\,135344\,B,  HD\,100546, AK\,Sco, and HD\,100453  present spirals that can be induced by the presence of a companion. For MWC\,758, \citet{2015ApJ...809L...5D} predict the presence of a $\sim 10$ M$_{Jup}$ companion at 160 au, which could have shaped the disk. Then, \citet{2018A&A...611A..74R} present a companion candidate at 20 au, and their contrast curves rule out companions larger than 5 \MJup beyond the spirals ($\sim$ 0\farcs6), assuming hot-start models. For HD\,135344\,B, similarly, \citet{2016ApJ...832..178V} and \citet{2019MNRAS.482.3609H}, connect the presence of the spiral features of the disk with the dynamical interaction with companions. For HD\,100546, protoplanet candidates were actually found: \citet{2015ApJ...807...64Q} confirm a candidate observed in different bands ($L^{\prime}$ and $M^{\prime}$). A second candidate was found by \citet{2015ApJ...814L..27C} using the Gemini Planet Imager \citep[GPI,][]{2006SPIE.6272E..0LM} instrument. Recently, both \citet{2017AJ....153..244R} and \citet{2018A&A...619A.160S} suggest that the candidates are more likely disk features.  \citet{2017A&A...608A.104M} found a bar-like structure in the polarized H$\alpha$ emission around the star. For AK\,Sco, \citet{2016ApJ...816L...1J} found two spiral arms in scattered light, which can be induced by the presence of a companion. For the spiral structures around HD\,100453, \citet{2016ApJ...816L..12D} prove, with dynamical simulations, that they are induced by the stellar object detected in NIR light at a projected distance of $\sim$ 119 au.  For the stars HD\,142527 \citep{2014ApJ...781L..30C} and HD\,98800\,BaBb \citep{2007ApJ...664.1176F}, for which accreting companions are found, we refer the reader to the dedicated Sections~\ref{s:HD142527} and \ref{s:HD98800}, respectively.

Some TDs present gaps and rings that can be carved by planets: around V4046\,Sgr\,AB, rings are found at 14 au and 29 au \citep{2015ApJ...803L..10R} as well as 40 au \citep{2018ApJ...863...44A}. The disk around HD\,169142 has a gap at 23 au \citep{2012ApJ...752..143H} and in scattered light at 40 au \citep{2017ApJ...850...52P}. Around this star, a protoplanet candidate was claimed by \citet{2014ApJ...792L..23R} and \citet{2014ApJ...792L..22B}, even if this detection has since been disputed \citep{2018MNRAS.473.1774L}.  DoAr\,44's disk has a gap from 5 au to 32 au \citep{2018MNRAS.477.5104C}; in scattered light it resembles a scaled down version of HD\,142527 \citep{2018ApJ...863...44A}. {The presence of an inner warp in this disk may be caused by a planet \citep{2018MNRAS.477.5104C}. } The TD around HD\,97048 has gaps at 34, 79, 140, and 179 au, and four rings \citep{2016A&A...595A.112G}. Very recently, \citet{2019NatAs.tmp..419P}, presented the detection of a doppler kink in the gas flow of the disk. This Keplerian deviation can be explained by the presence of a planet that perturbs the gas flow \citep{Perez2015}. The kink is located exactly in one of the gaps, which is seen in both the ALMA continuum and in the scattered light images. For HD\,141569\,A, \citet{2016A&A...590L...7P} found ringlets at 47 au, 64 au, and 93 au. The UX\,Tau\,A disk has an inner cavity at 56 au \citep{2007ApJ...670L.135E}. PDS\,66 has an inner cavity up to 15 au \citep{2013A&A...552A..88G,2016ApJ...818L..15W}.




\subsection{Observations}

This survey has been carried out with the instrument SPHERE \citep{2019A&A...631A.155B}. SPHERE is a planet finder at the ESO's VLT that is equipped with an extreme adaptive optics system with a $41\times41$ actuators wavefront control, pupil stabilization, and differential tip-tilt control \citep{2014SPIE.9148E..0OP}. The instrument has three science subsystems: the infrared dual-band imager and spectrograph \citep[IRDIS;][]{Do08}, an integral field spectrograph \citep[IFS;][]{Cl08}, and the Zimpol rapid-switching imaging polarimeter \citep[ZIMPOL;][]{Th08,2018A&A...619A...9S}. The latter, which is the only subsystem in visible light, has been used for the survey. The observing strategy was to take simultaneous images in the H$\alpha$ narrow band filter (N\_Ha; $\lambda_c$ = 656.9 nm, $\Delta\lambda$ = 1 nm) and in H$\alpha$ continuum (Cont\_Ha; $\lambda_c$ =  644.9 nm, $\Delta\lambda$ = 4.1 nm). We refer the reader to \cite{2018A&A...619A...9S} for more details about the filters. All the observations were taken in pupil stabilized mode in order to take advantage of the speckles suppression technique of the angular differential imaging \citep[ADI;][]{2006ApJ...641..556M}.

The observations (ESO programs 099.C-0453, 0100.C-0193, 0101.C-0461, 0102.C-0138, PI: Zurlo) were carried out in service mode, with the exception of the target DoAr\,44 for which visitor mode was required during the last four ESO observing periods. To complete the sample for all the TDs that are closer than 200 pc and observable with SPHERE, we included the objects MWC\,758, HD\,135344B, and HD\,142527 (programs 096.C-0267, PI: Huelamo, 096.C-0248, SPHERE GTO). The observations and analysis of the dataset of MWC\,758 are presented in \citet{2018A&A...613L...5H}. For this object, the broad band H$\alpha$ filter (B\_Ha; $\lambda_c$ = 655.6 nm, $\Delta\lambda$ = 3.35 nm) was used. This same dataset, together with the datasets of the objects HD\,135344B and HD\,142527 are analysed in \citet{2019A&A...622A.156C}. The different observing dates for each target, together with the total field of view rotation and the mean value of the seeing, are listed in Table~\ref{t:obs}. In general, the conditions were stable and the seeing was below 0.8\as. The total execution time per target was in between 1h and 1h30min.   

\begin{center}
  \begin{table*}
    \begin{minipage}{1\textwidth}
    \caption{Summary table of the observations of the survey.} 
    \label{t:obs}
    \renewcommand{\footnoterule}{} 
\centering
    \begin{tabular}{lcccccccc}
\hline
 Name & {V mag} & Obs date & DIT\footnote{Detector Integration Time}xNDIT\footnote{Number of frames per dithering position}& Rot (deg)  & Seeing (\as) &Airmass  \\
\hline

MWC 758$^{*}$ & 8.3 &  30/12/2015  & 60x10   & 58  &  1.3 & 1.6   \\
HD 135344\,B$^{*}$  & 8.7    &  31/03/2016 & 50x15  &  72  & 0.6 & 1.0 \\
HD 142527$^{*}$ & 8.3 & 31/03/2016 & 30x10  & 49   & 0.7 & 1.0 \\
 HD 100546 & 6.3 &24/03/2016 & 120x5  & 57   &  2.8  & 1.5         \\
  & & 16/03/2019 & 9x20 & 27  & 0.8 & 1.4 \\
HD 98800\,B & 10.1 &  08/01/2017 & 11x10  & 1  & 0.4 & 1.0\\
 & & 12/01/2017 & 11x10 &  1  & 0.6 & 1.0 \\
AK Sco &9.2 & 20/05/2017 & 40x4 &  39 &  0.5 & 1.0  \\ 
V4046 Sgr AB &10.7 &  20/05/2017 &  60x4 & 99  & 0.5 & 1.0 \\
& & 14/07/2017 & 15x5 &  39  & 0.6 & 1.0 \\  
HD 169142 & 8  & 28/05/2017& 8x5  & 125  & 0.8  & 1.0 \\  
& & 17/05/2018  &9x17 &   84  & 0.4  & 1.0 \\
DoAr 44 & 11.7 &  29/06/2017 & 60x2 & 3 & 0.9 & 1.0 \\ 
HD 97048 & 8.5 & 06/02/2018 &7x15 &  23  & 0.3 & 1.7 \\
  &  & 22/02/2019 &11x20 & 23 & 0.7 & 1.7  \\
HD 141569 A & 7.1 &  22/03/2018 & 8x18 &  29  & 0.7 & 1.0 \\  
  & & 17/05/2018 & 11x20  & 43 & 0.5 & 1.0  \\
UX Tau\,A & 11.1 & 26/11/2018 & 110x2  & 24 &  0.4  & 1.4 \\
HD 100453 & 7.8 & 20/01/2019 & 11x20 & 36  & 0.5  & 1.2\\

PDS 66 & 10.4 & 22/03/2019 & 24x10    &  34 & 0.5 & 1.5  \\
 \hline
    \end{tabular}
   \end{minipage}
    \end{table*}
\end{center}

\section{Data reduction}
\label{sec:red}

The raw data were preprocessed using the ZIMPOL pipeline, an {\it IDL} routine developed at ETH Zurich. The pipeline produces bias-subtracted, flat-fielded, remapped images for each of the two filters. We refer the reader to \citet{2019A&A...622A.156C}, for further details. We used a python routine to recenter the images, with a
Gaussian two-dimensional fitting and to create two different data cubes, one for the H$\alpha$ filter and one for the continuum. The python package {\it VIP} \citep{2017AJ....154....7G} is used in this analysis. To perform the spectral differential imaging \citep[SDI;][]{2007ApJ...660..770L} technique each image in the continuum filter was subtracted to the H$\alpha$ filter after being rescaled and normalized for the central point spread function (PSF).

{More specifically, we first rescaled the continuum images. Additionally, in order to normalize the two PSFs, we used a minimum $\chi$-square method, which minimizes the standard deviation inside a circle centered in the center of the image and with a radius of 5$\times$FWHM, after the subtraction of the two images. In this way, we are able to take the center of the PSF into account, in addition to its wings and the bright central speckles.   }

We then performed a principal component analysis (PCA) reduction to apply the ADI method with a different number of components, depending on the total number of frames of each observation. In most of the cases, a frame selection rejecting low-quality frames {(with a maximum of the 25\% of the total number)} was applied. We refer to angular spectral differential imaging (ASDI) reduction as the combination of angular and spectral differential imaging. To align the images, we applied a true north correction of 134 $\pm$ 0.5 deg as in \citet{2019A&A...622A.156C}. 

\begin{figure}
\begin{center}
\includegraphics[width=0.5\textwidth]{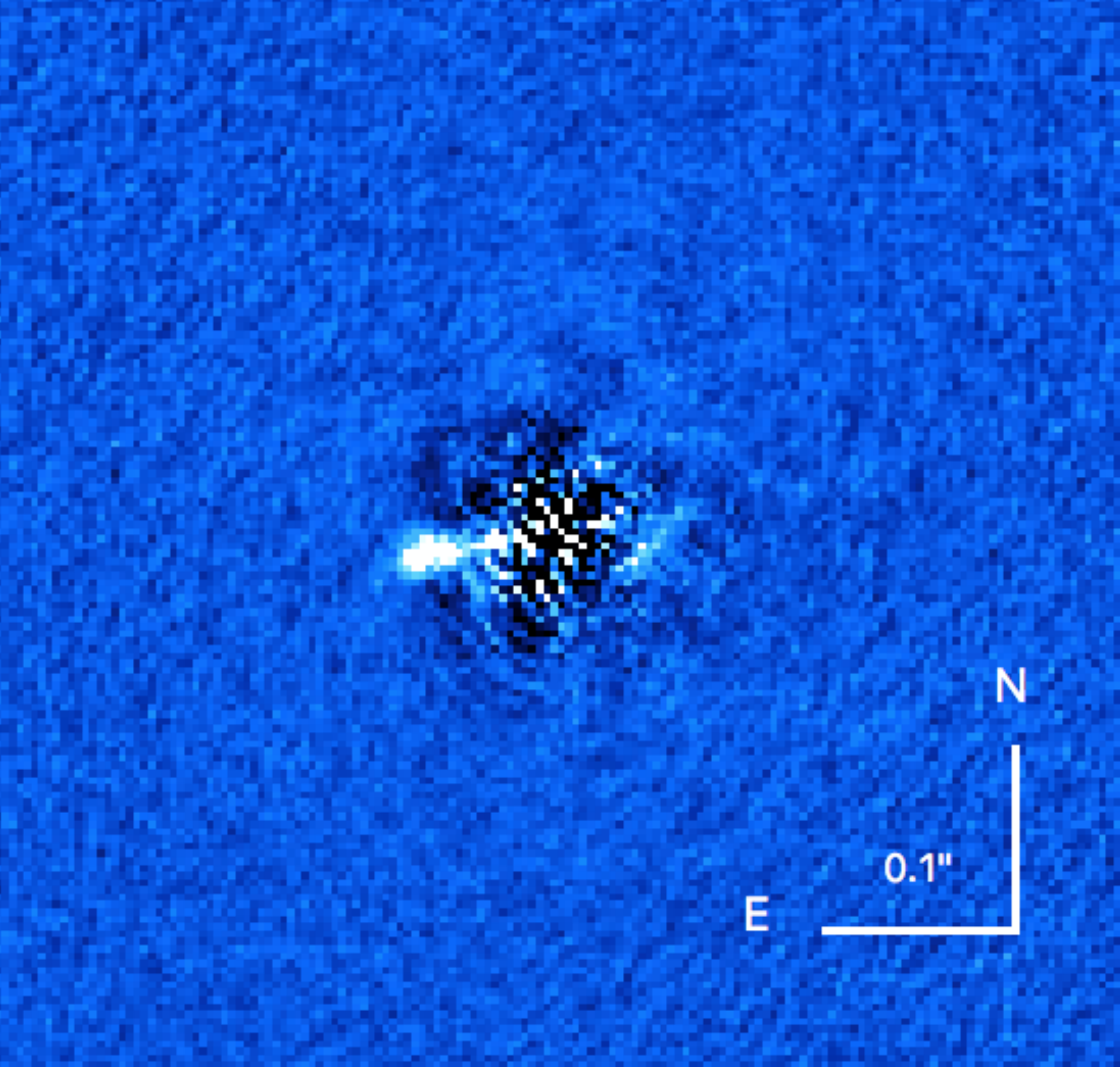}
\caption{ASDI image of the system HD\,142527. The stellar accreting companion is visible in the East. The image shows the very close vicinity of the stellar companion. }
\label{f:HD142527fits}
\end{center}
\end{figure}

\begin{figure}
\begin{center}
\includegraphics[width=0.5\textwidth]{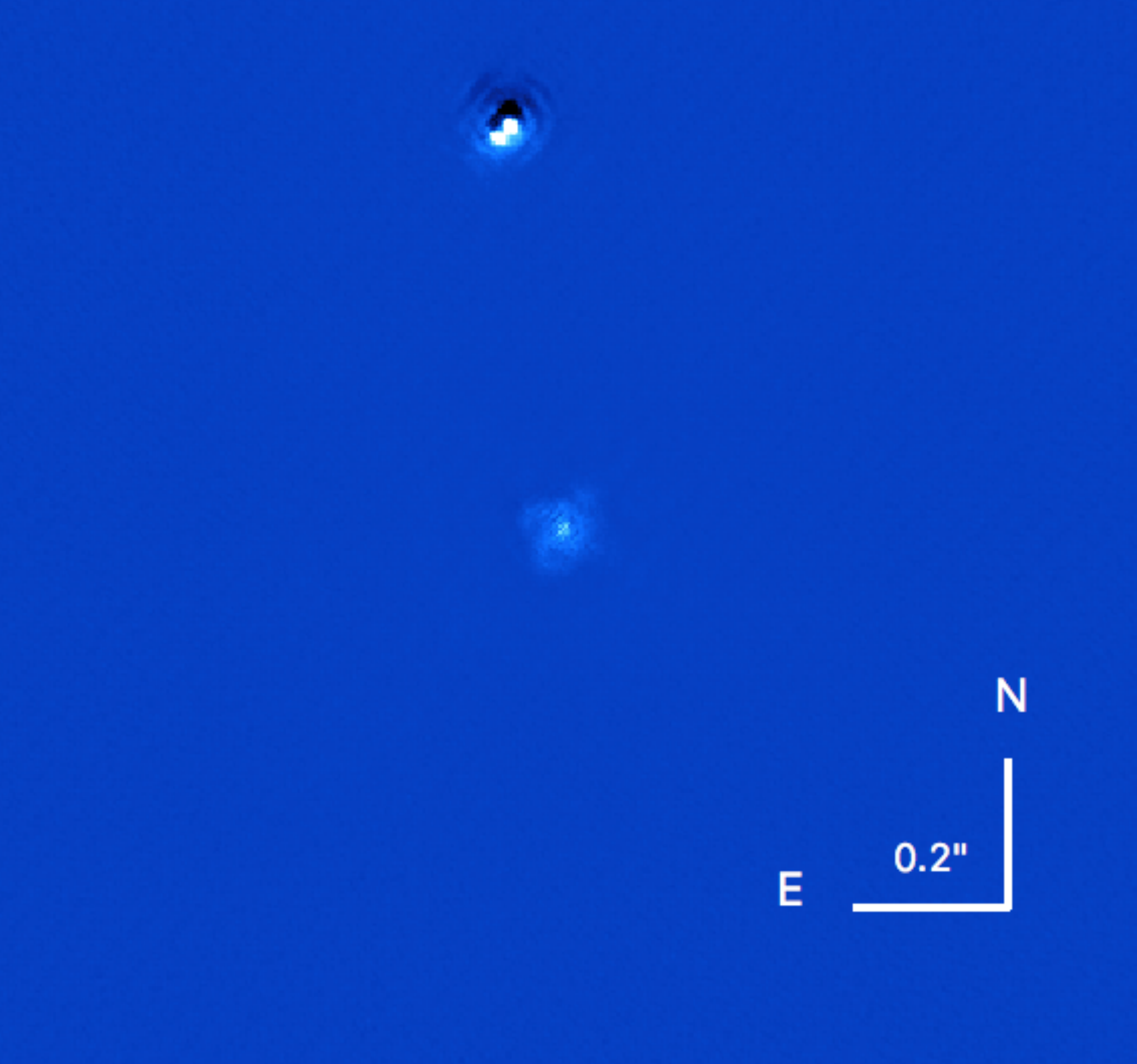}
\caption{ASDI image of the system HD\,98800\,BaBb. The system is a quadruple: two spectroscopic binaries at the center of the detector and two stellar companions (one accreting) are visible in the north, which are resolved for the first time in the visible. }
\label{f:HD98800Bfits}
\end{center}
\end{figure}

A couple of examples of the final ASDI PCA image are shown in Fig.~\ref{f:HD142527fits} for the object HD\,142527\,B and Fig.~\ref{f:HD98800Bfits} for the quadruple system around HD\,98800. The two stellar companions of the spectroscopic binary were resolved at these wavelengths for the first time.  

\section{Analysis and results}
\label{sec:anal}

No significant detection of point sources has been recovered from the data reduction, apart from the already known stellar companion of HD\,142527, which is presented in Sec.~\ref{s:HD142527}, and the two stellar companions of HD\,98800\,Ba, in Sec.~\ref{s:HD98800}. On the other hand, the contrast curves produced reflect the good conditions and exquisite performance of the instrument. We refer the reader to \citet{2014A&A...572A..85Z,2016A&A...587A..57Z} for a detailed explanation of how we calculated the contrast curves.

\begin{figure}
\begin{center}
\includegraphics[width=0.5\textwidth]{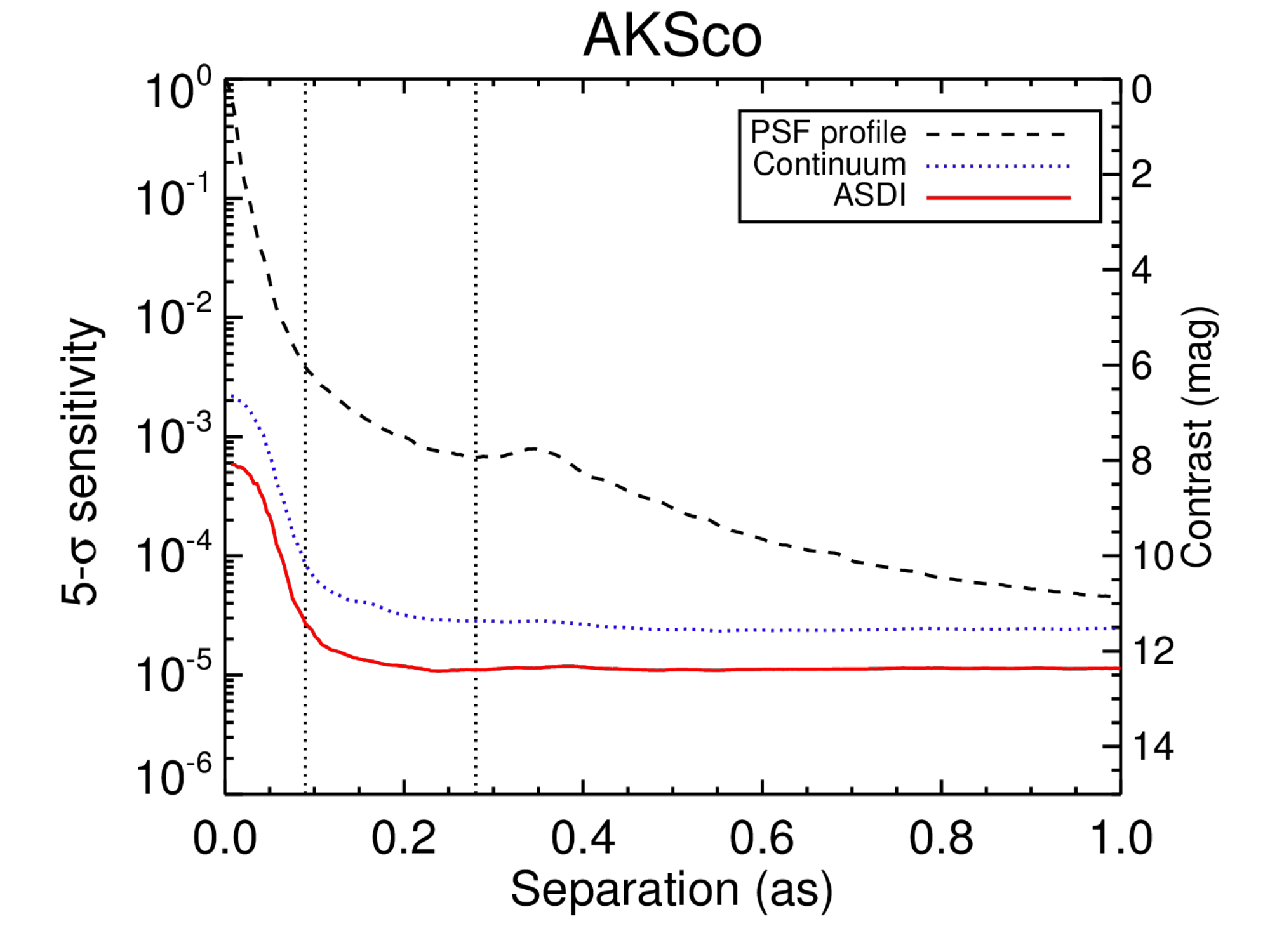}
\caption{Contrast plot for the object AK\,Sco. In a black solid line, the profile of the PSF is shown, blue represents the H$\alpha$ continuum filter after being processed with PCA and ADI, and red is the ASDI contrast. The vertical dotted lines represent the extension of the spirals \citep{2016ApJ...816L...1J}.   }
\label{f:k2}
\end{center}
\end{figure}

\begin{figure}
\begin{center}
\includegraphics[width=0.5\textwidth]{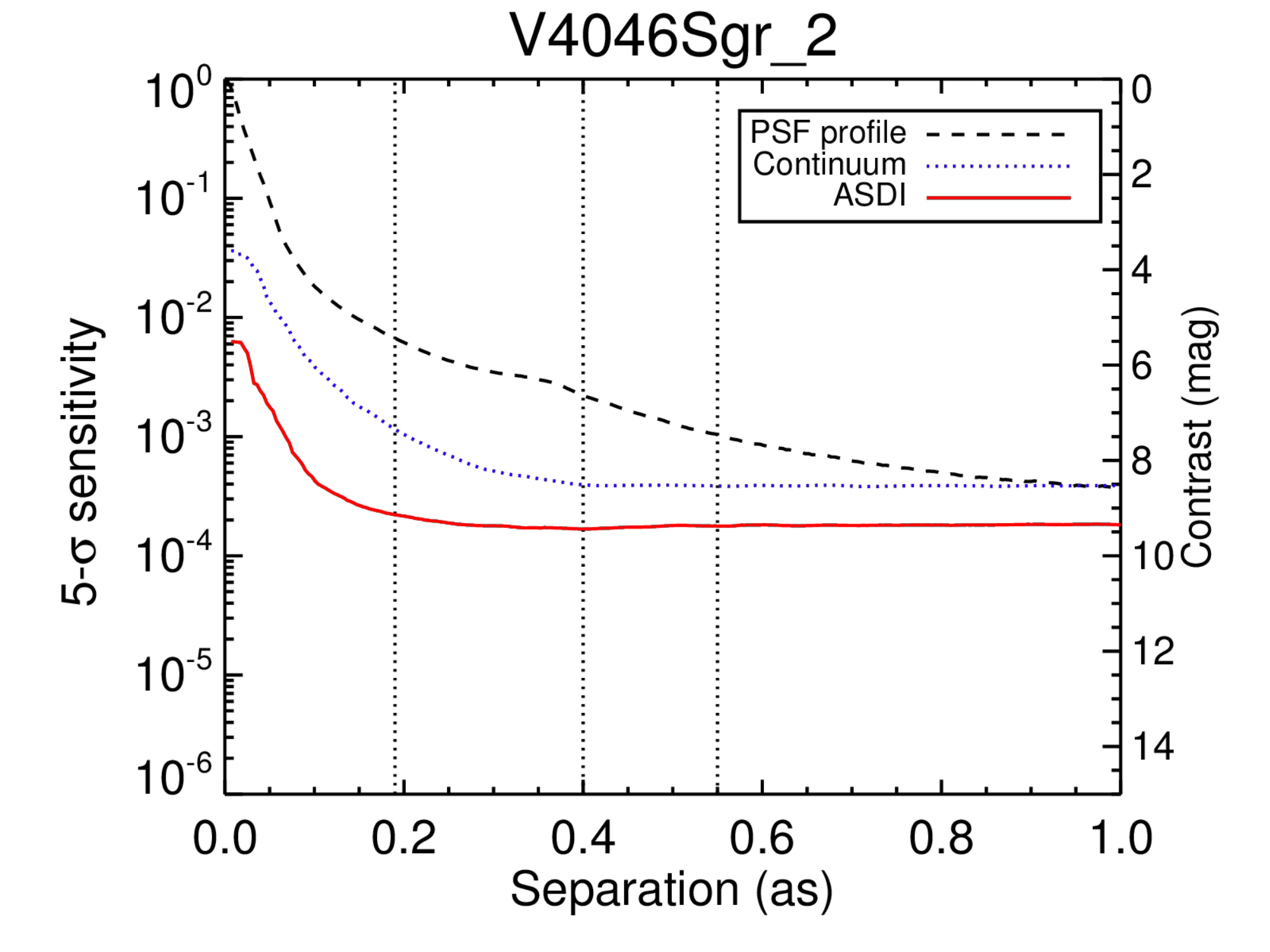}
\caption{Same as Fig.~\ref{f:k2}, but for the object V4046\,Sgr. The vertical dotted lines represent the locations of the rings detected by \citet{2015ApJ...803L..10R} and \citet{2018ApJ...863...44A}.  }
\label{f:V4046Sgr_2}
\end{center}
\end{figure}

\begin{figure}
\begin{center}
\includegraphics[width=0.5\textwidth]{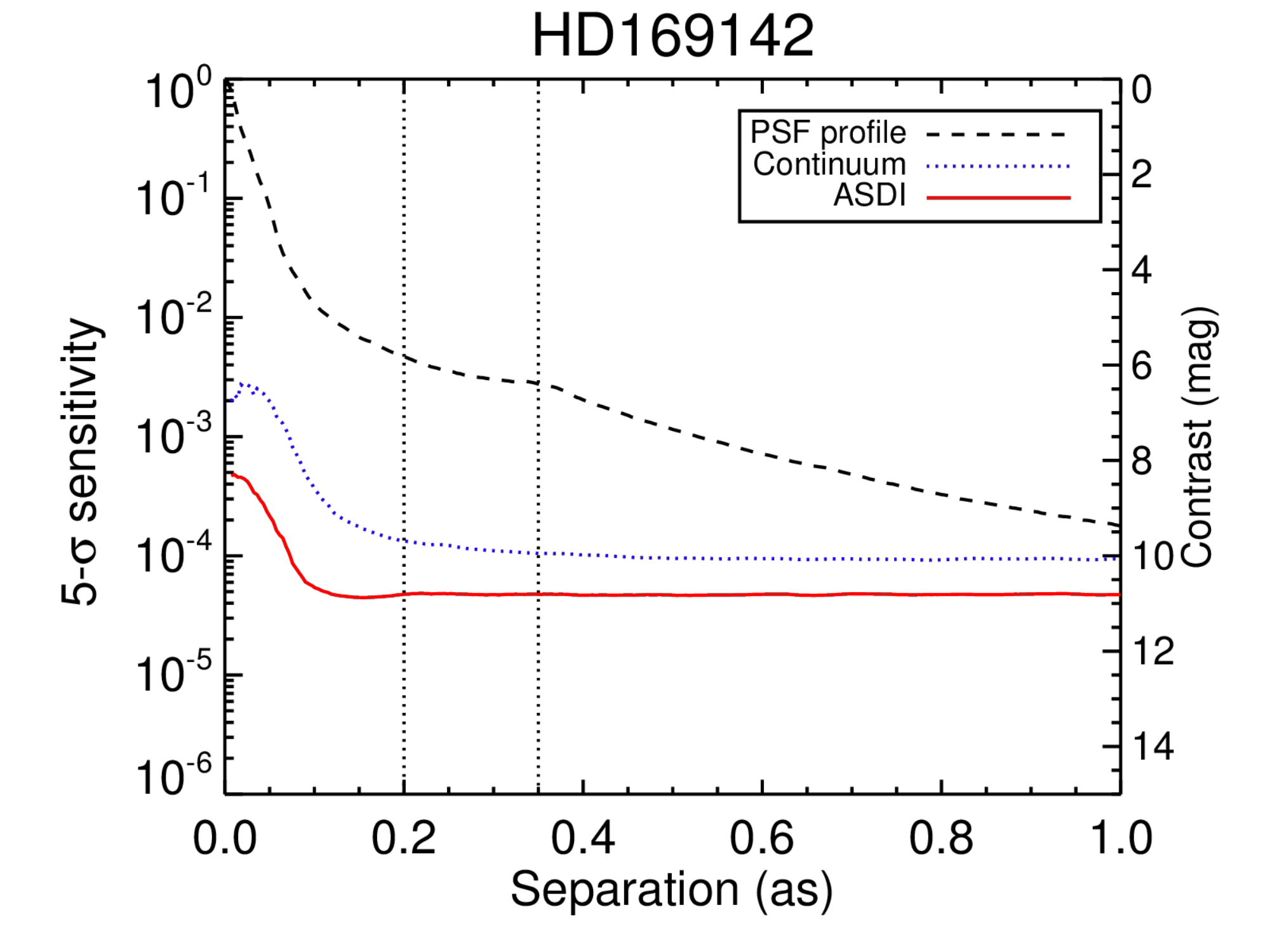}
\caption{Same as Fig.~\ref{f:k2}, but for the object HD\,169142. The vertical dotted lines represent the location of the gaps detected by \citet{2012ApJ...752..143H} and \citet{2017ApJ...850...52P}. }
\label{f:HD169142}
\end{center}
\end{figure}

\begin{figure}
\begin{center}
\includegraphics[width=0.5\textwidth]{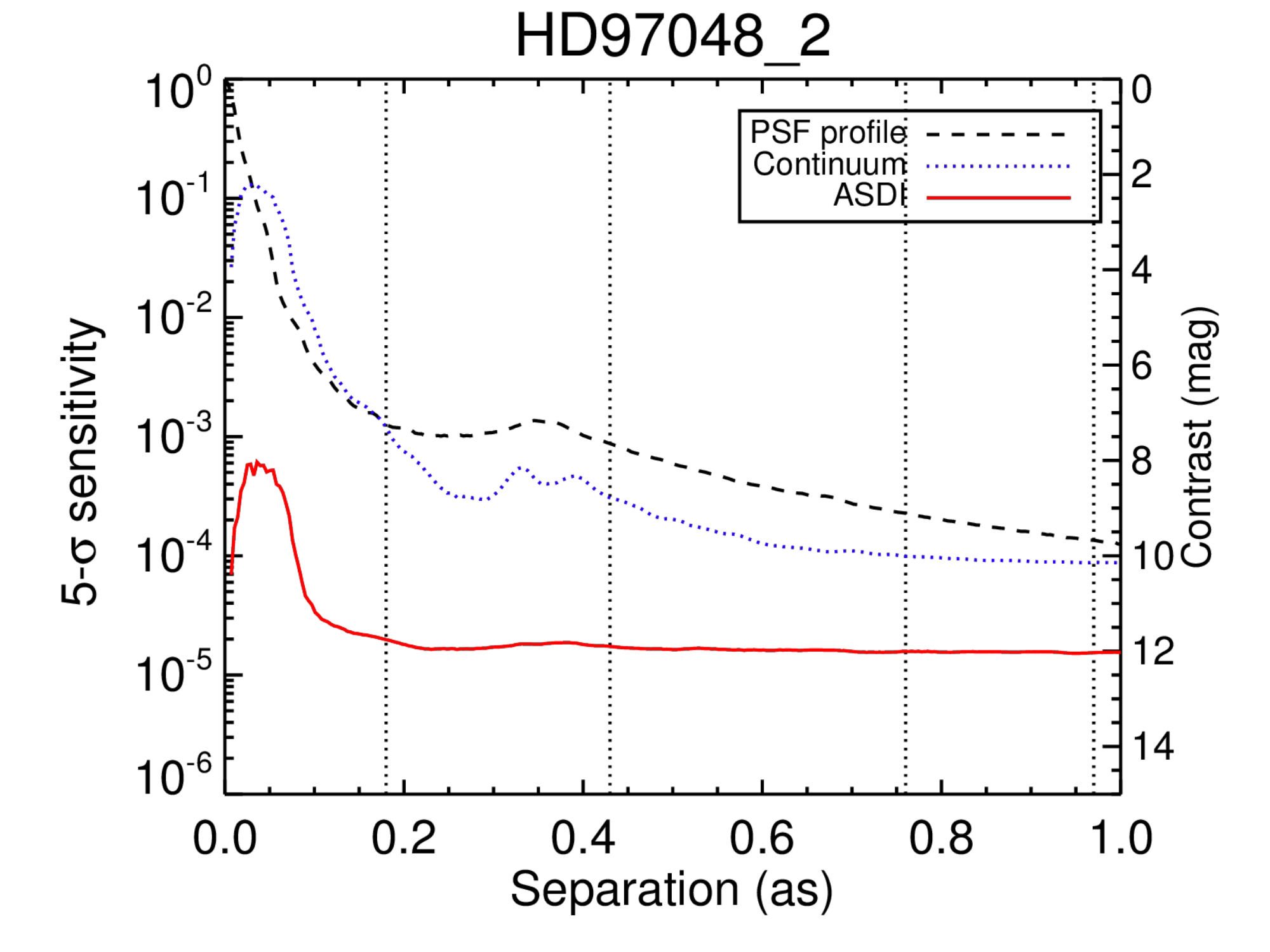}
\caption{Same as Fig.~\ref{f:k2}, but for the object HD\,97048. The vertical dotted lines represent the location of the gaps detected by \citet{2016A&A...595A.112G}. }
\label{f:HD97048}
\end{center}
\end{figure}

\begin{figure}
\begin{center}
\includegraphics[width=0.5\textwidth]{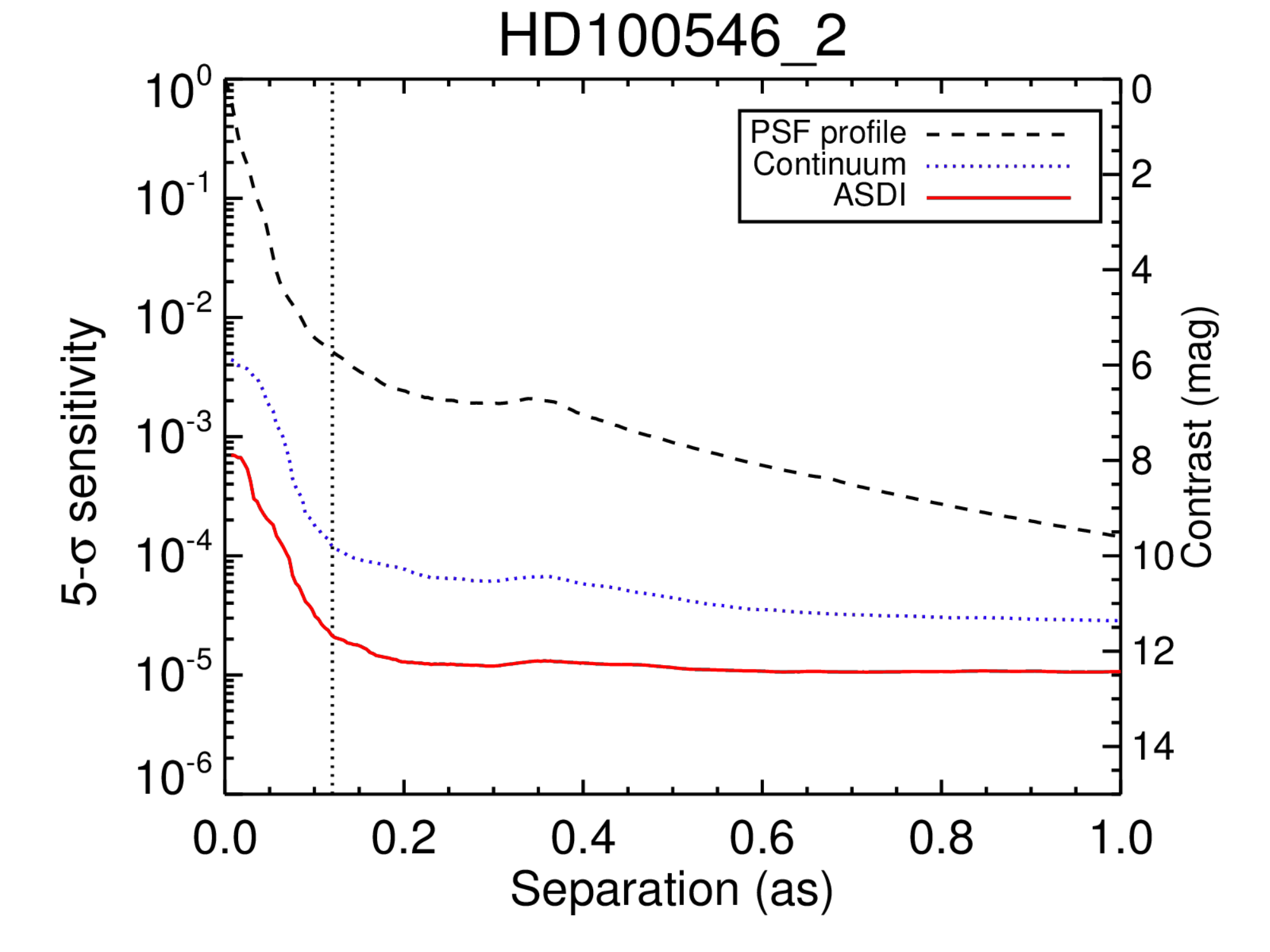}
\caption{Same as Fig.~\ref{f:k2}, but for the object HD\,100546. The dotted vertical line is located at the position of the inner gap \citep{2015ApJ...807...64Q}.}
\label{f:HD100546}
\end{center}
\end{figure}

\begin{figure}
\begin{center}
\includegraphics[width=0.5\textwidth]{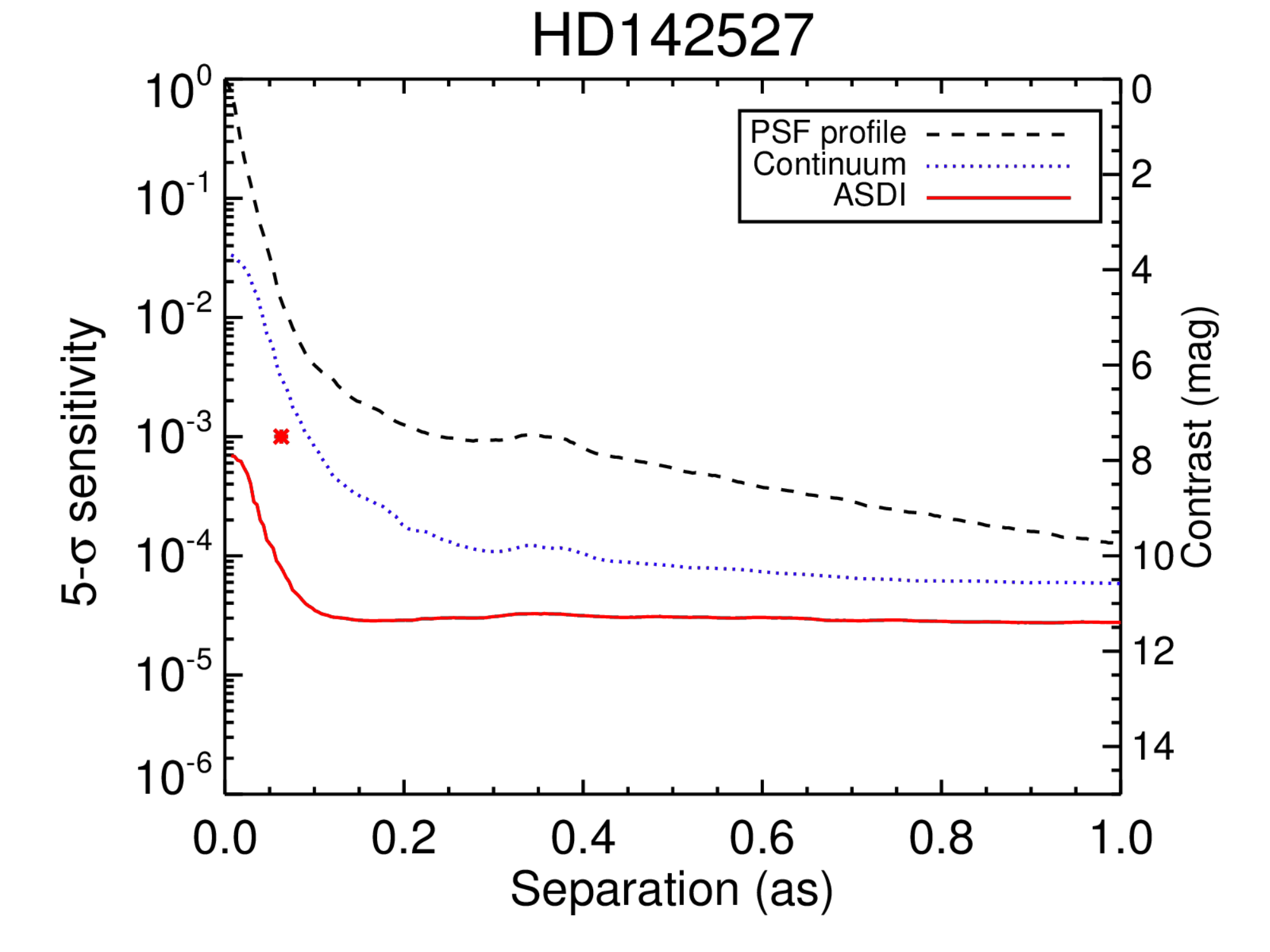}
\caption{Same as Fig.~\ref{f:k2}, but for the object HD\,142527. {The red point represents the detected stellar companion.} }
\label{f:HD142527}
\end{center}
\end{figure}

\begin{figure}
\begin{center}
\includegraphics[width=0.5\textwidth]{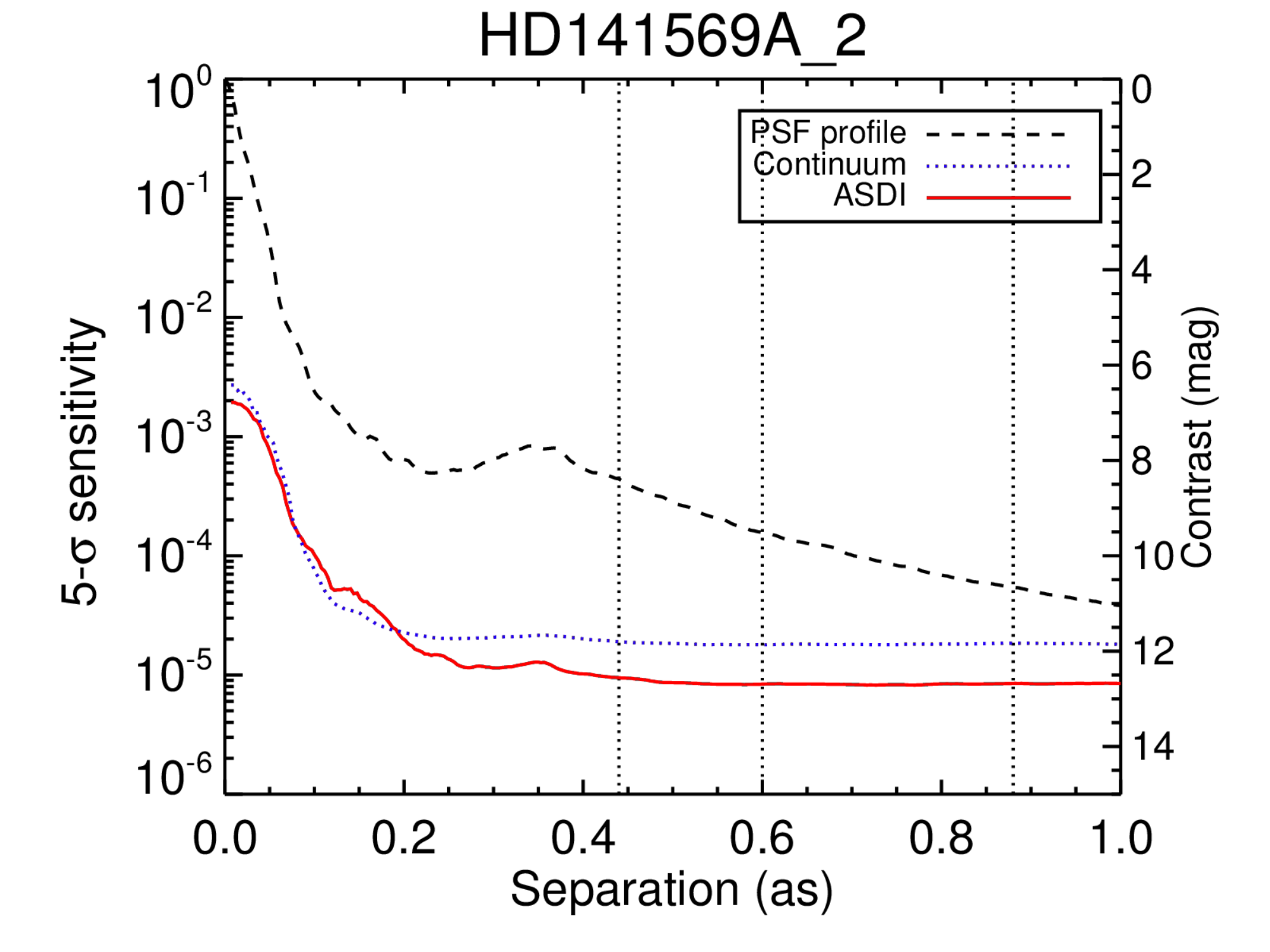}
\caption{Same as Fig.~\ref{f:k2}, but for the object HD\,141569A. The vertical dotted lines represent the location of the ringlets detected by \citet{2016A&A...590L...7P}. }
\label{f:HD141569A}
\end{center}
\end{figure}

\begin{figure}
\begin{center}
\includegraphics[width=0.5\textwidth]{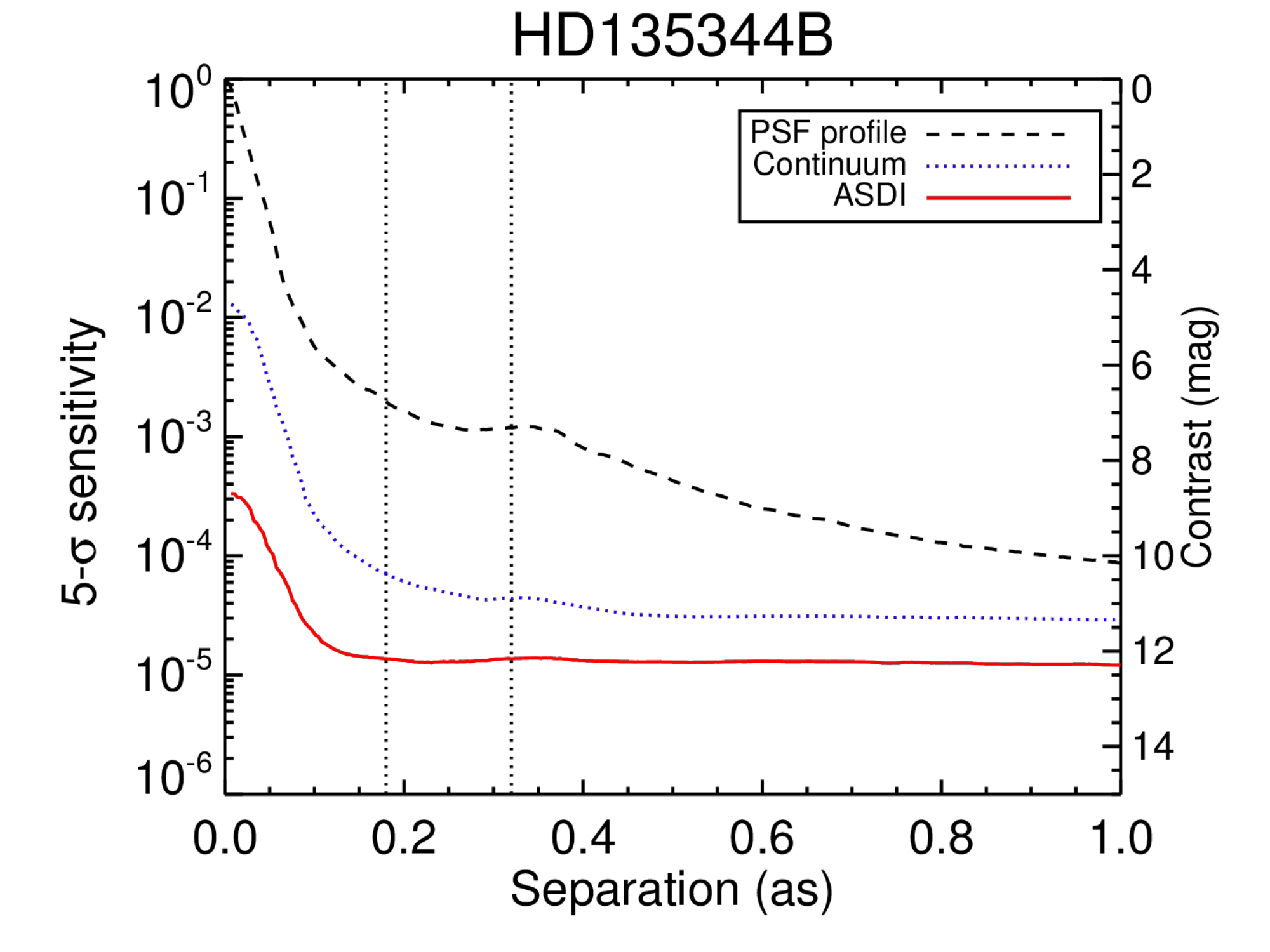}
\caption{Same as Fig.~\ref{f:k2}, but for the object HD\,135344B. The two dotted lines represent the extension of the spiral arms detected by \citet{2016ApJ...832..178V} and \citet{2019MNRAS.482.3609H}. }
\label{f:HD135344B}
\end{center}
\end{figure}

\begin{figure}
\begin{center}
\includegraphics[width=0.5\textwidth]{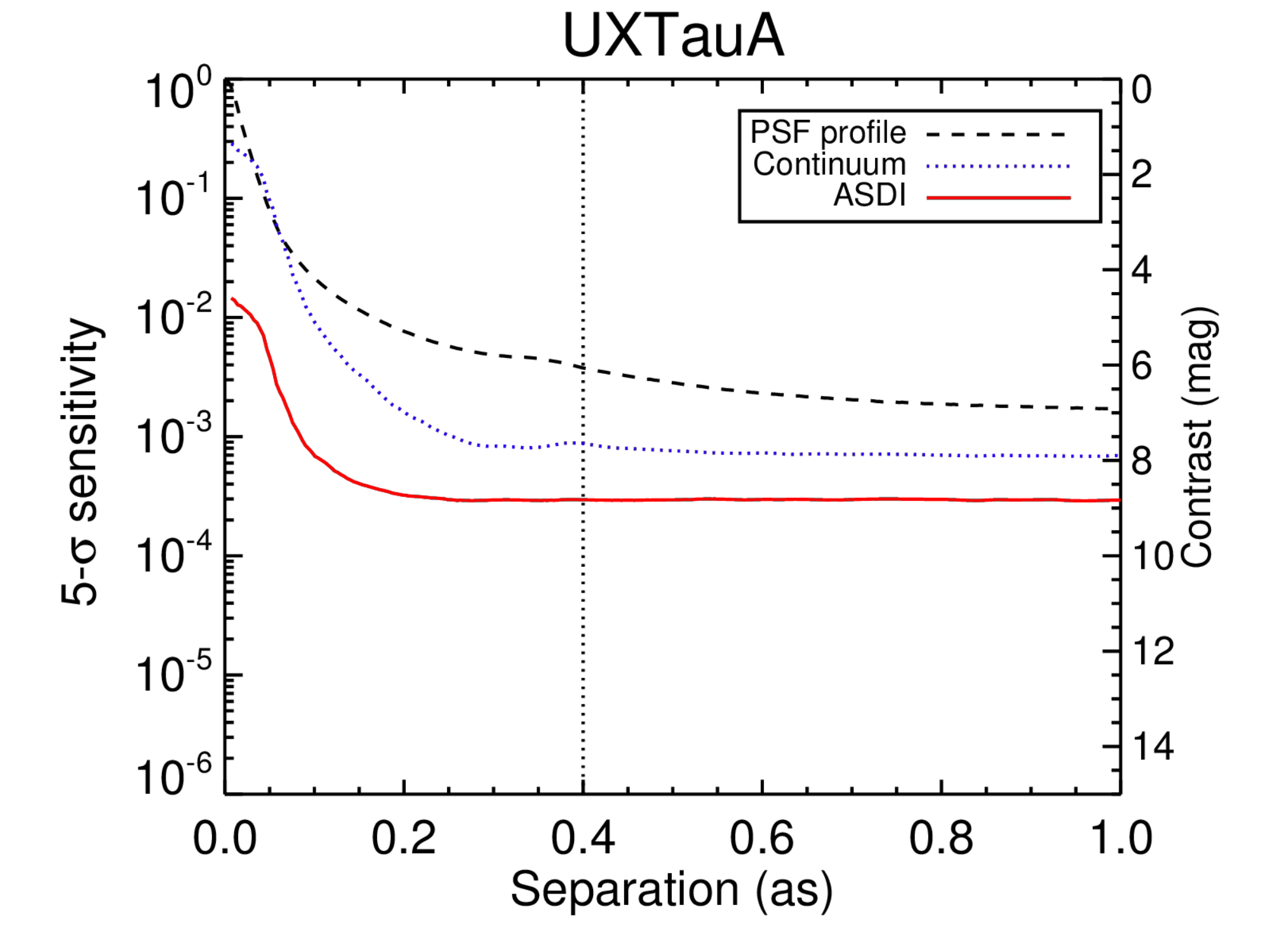}
\caption{Same as Fig.~\ref{f:k2}, but for the object UX\,TauA. The vertical dotted line represents the location of the gap \citep{2007ApJ...670L.135E}. }
\label{f:UXTauA}
\end{center}
\end{figure}

\begin{figure}
\begin{center}
\includegraphics[width=0.5\textwidth]{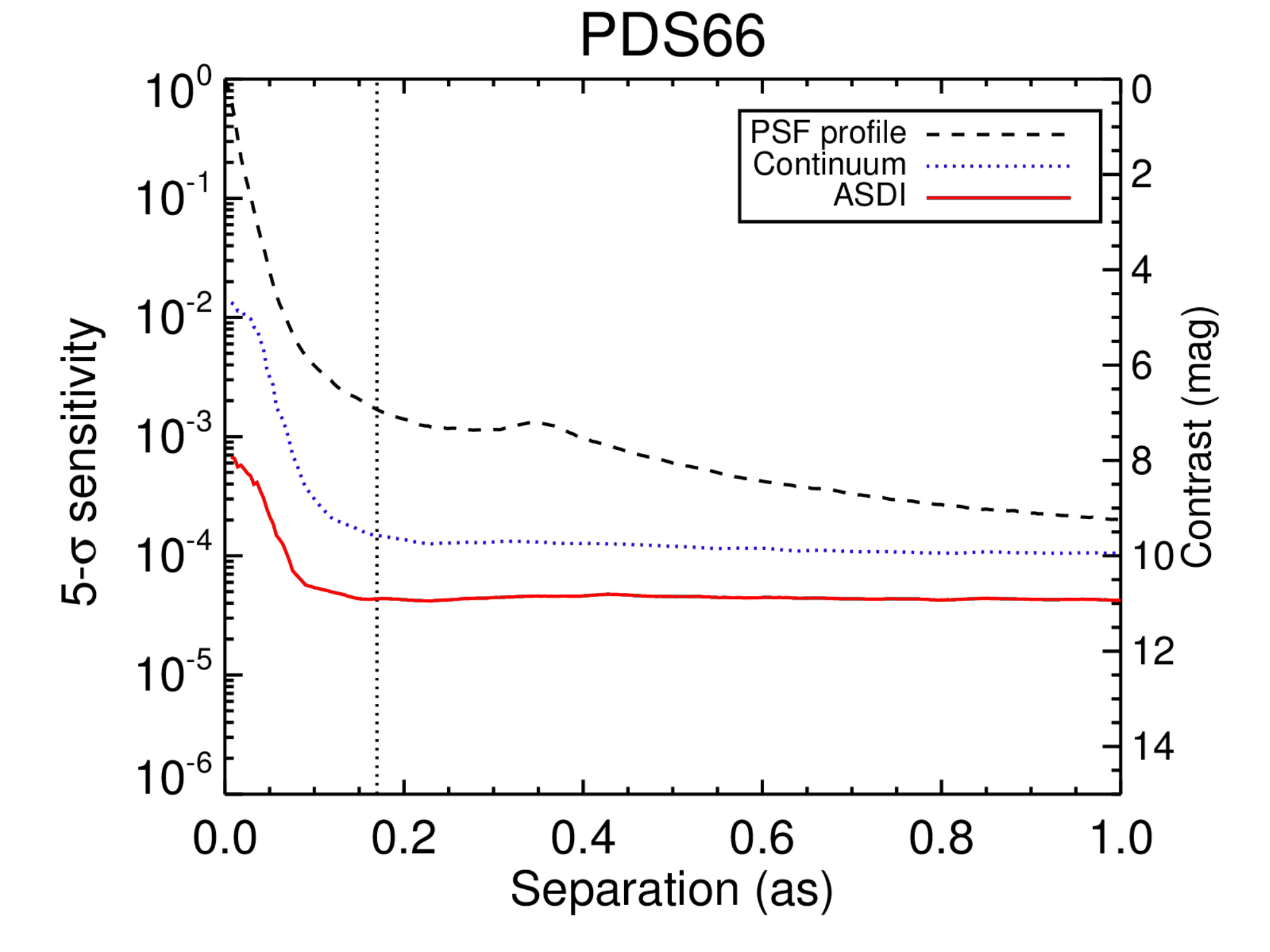}
\caption{Same as Fig.~\ref{f:k2}, but for the object PDS\,66. The vertical dotted line represents the location of the inner cavity \citep{2013A&A...552A..88G,2016ApJ...818L..15W}.}
\label{f:PDS66}
\end{center}
\end{figure}

\begin{figure}
\begin{center}
\includegraphics[width=0.5\textwidth]{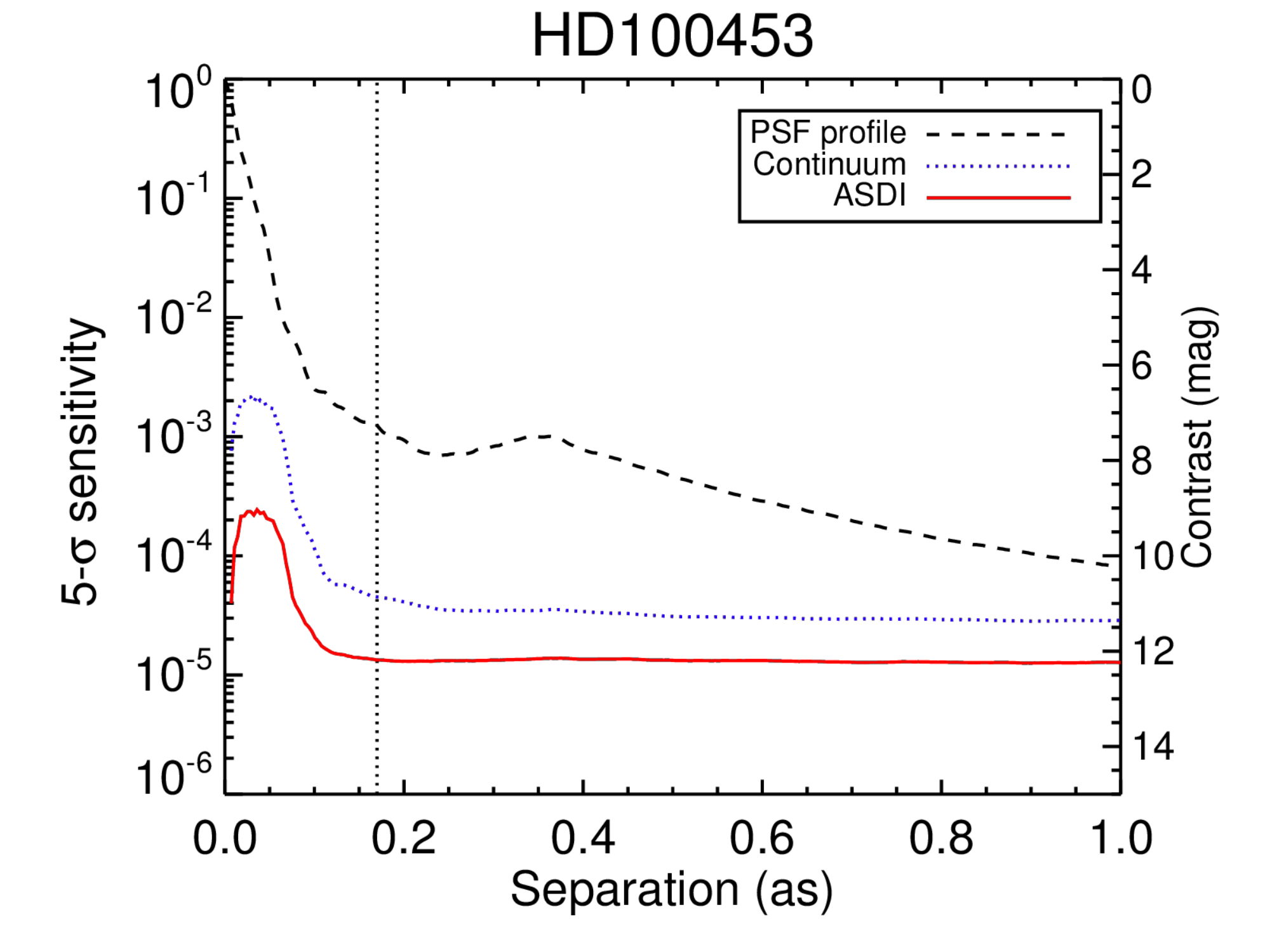}
\caption{Same as Fig.~\ref{f:k2}, but for the object HD\,100453. The vertical dotted line represents the inner part of the spirals detected by \citet{2017ApJ...835...38D}. }
\label{f:HD100453}
\end{center}
\end{figure}

In Figure~\ref{f:k2}, we present the contrast curves for the target AK\,Sco in addition to V4046Sgr\,A (Fig.~\ref{f:V4046Sgr_2}), HD\,169142 (Fig.~\ref{f:HD169142}), HD\,97048 (Fig.~\ref{f:HD97048}),  HD\,100546 (Fig.~\ref{f:HD100546}),  HD\,142527 (Fig.~\ref{f:HD142527}), HD\,141569A (Fig.~\ref{f:HD141569A}), HD\,135344B (Fig.~\ref{f:HD135344B}), UX\,TauA (Fig.~\ref{f:UXTauA}), PDS\,66 (Fig.~\ref{f:PDS66}), and HD\,100453 (Fig.~\ref{f:HD100453}). In the case of multiple datasets for the same object we selected and show here the best one. The reduction was performed for all the datasets. In general, we could reach a very deep contrast of $\sim$ 12 mag at $0\farcs2$ separation from the host star.

\begin{figure*}
\begin{center}
  \includegraphics[width=0.45\textwidth]{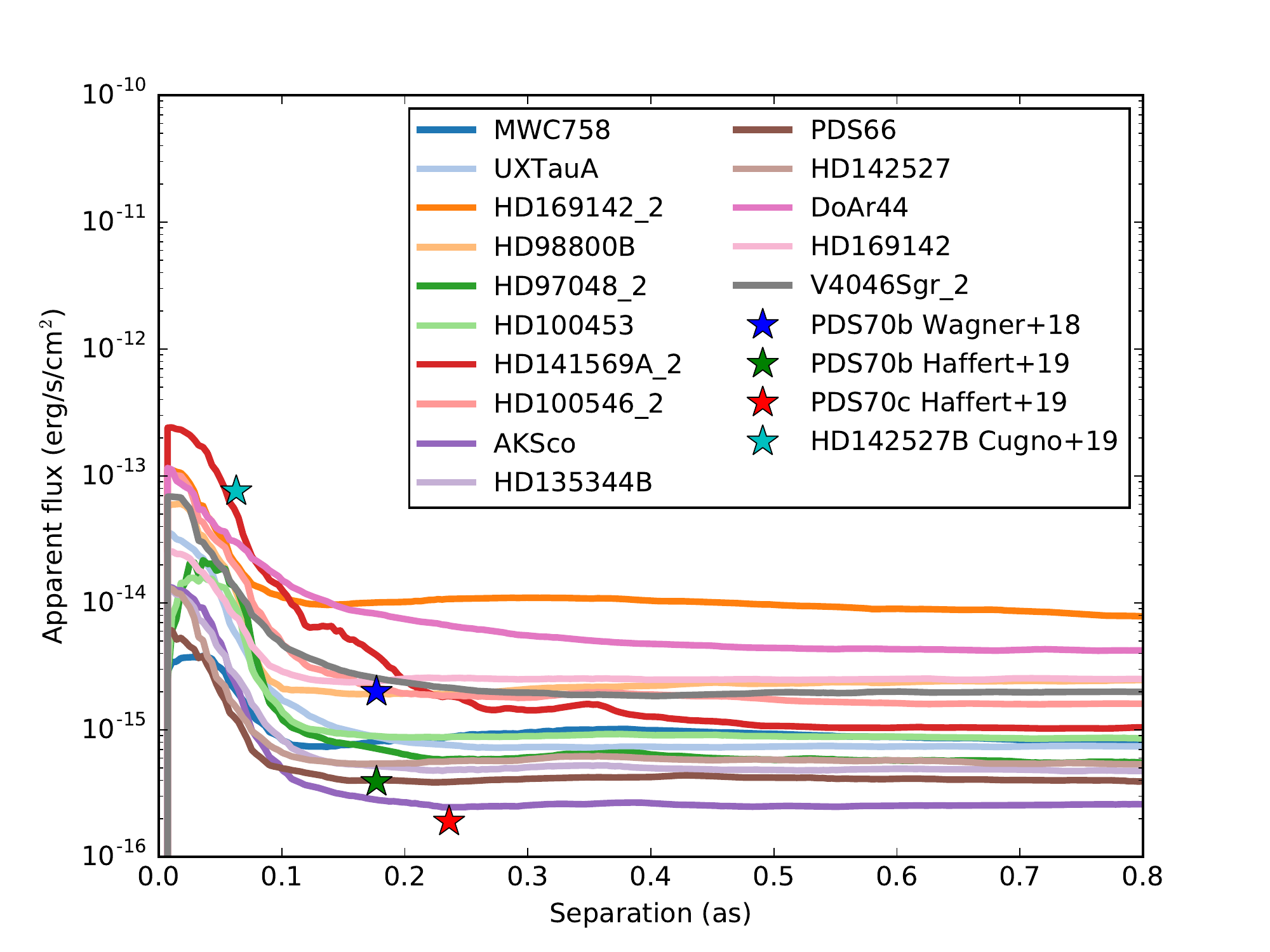}
  \includegraphics[width=0.45\textwidth]{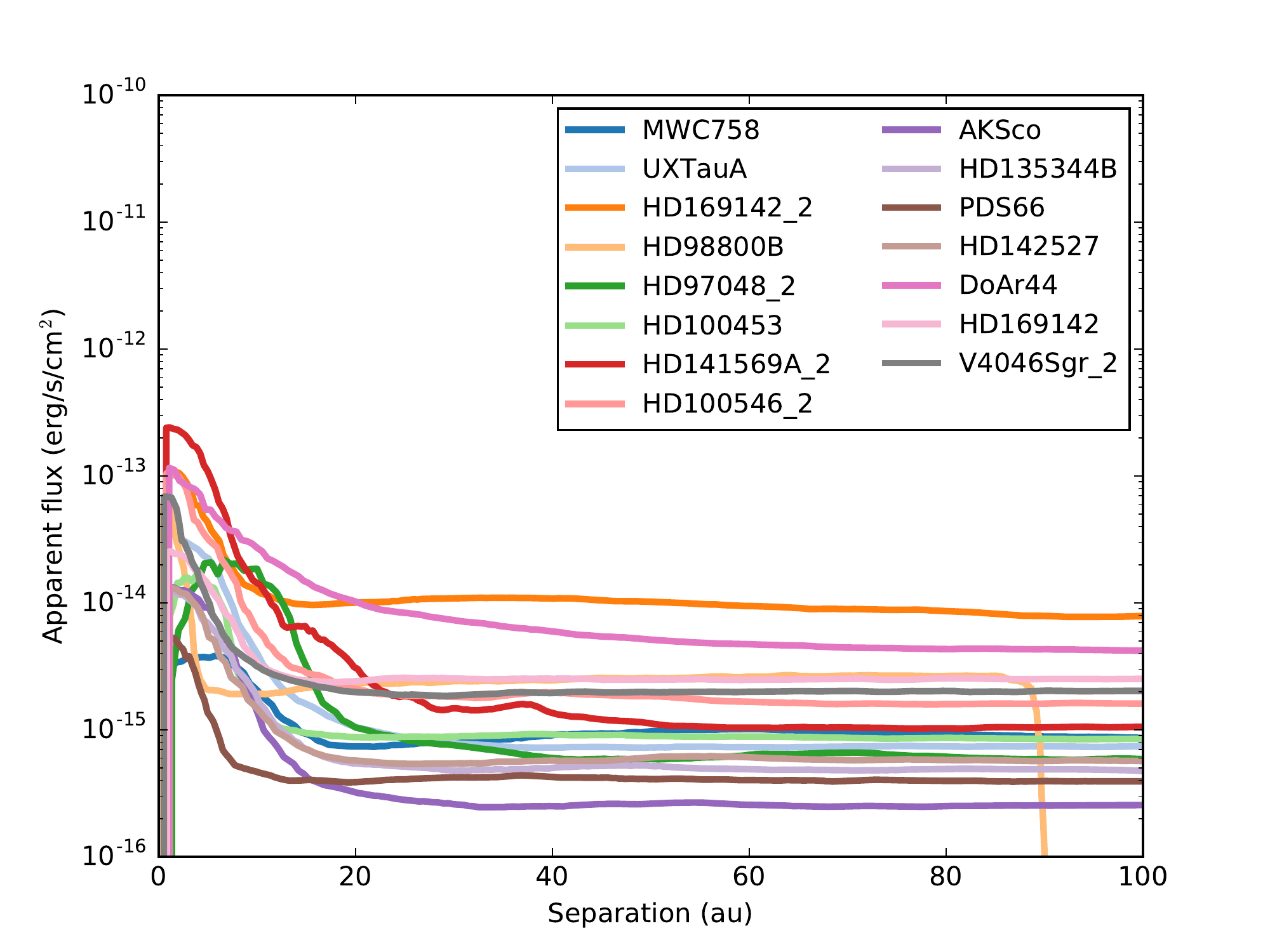}
\caption{Apparent H$\alpha$ line flux limits for all the targets of the sample versus angular separation (left) and physical separation (right). {The differences between our predictions and other cases from the literature are primarily related to the methods used to extrapolate the accretion luminosity from the H$\alpha$ line flux; as with SPHERE, the line width cannot be measured.} }
\label{f:flux}
\end{center}
\end{figure*}

\begin{figure*}
\begin{center}
  \includegraphics[width=0.45\textwidth]{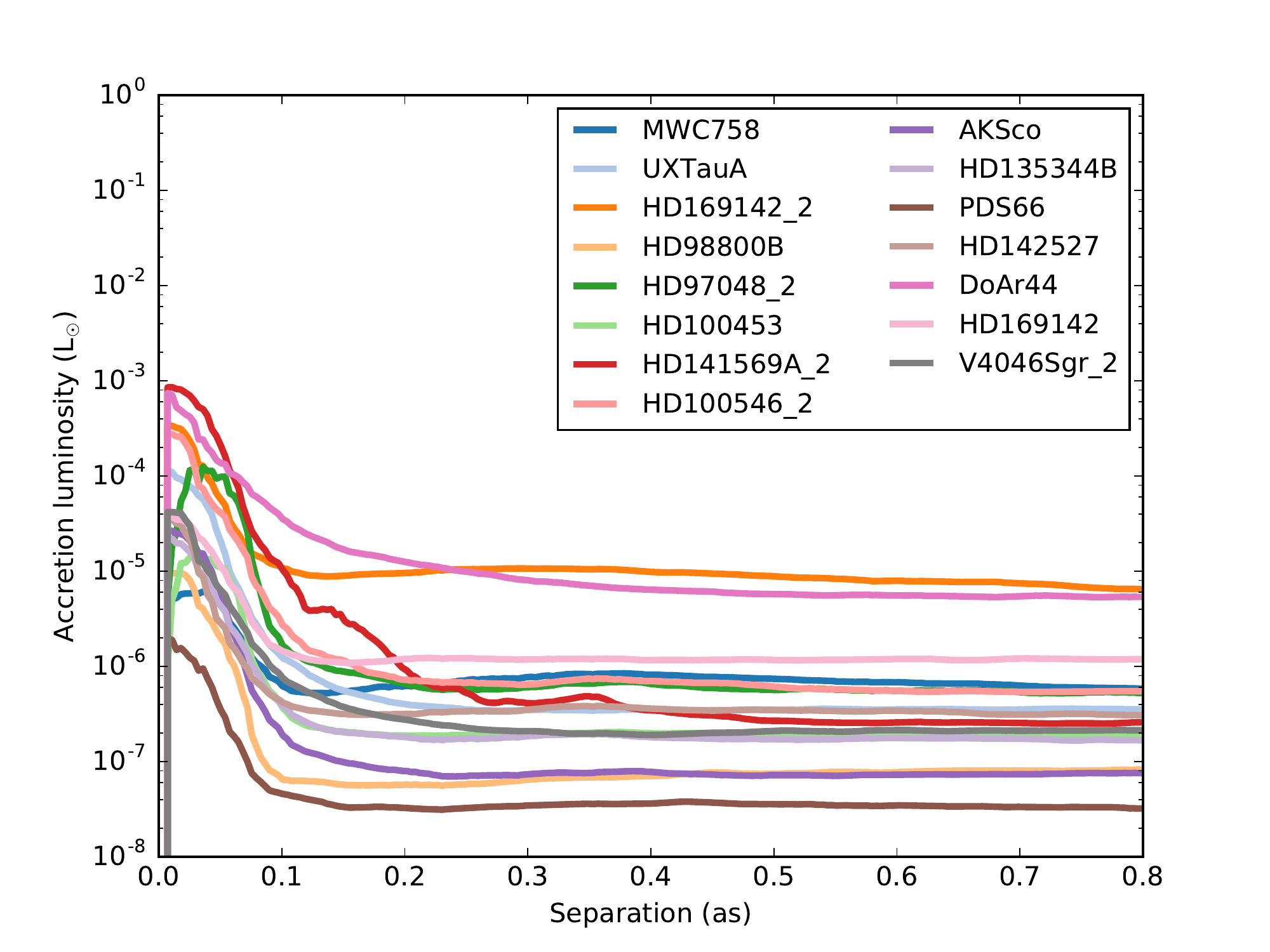}
  \includegraphics[width=0.45\textwidth]{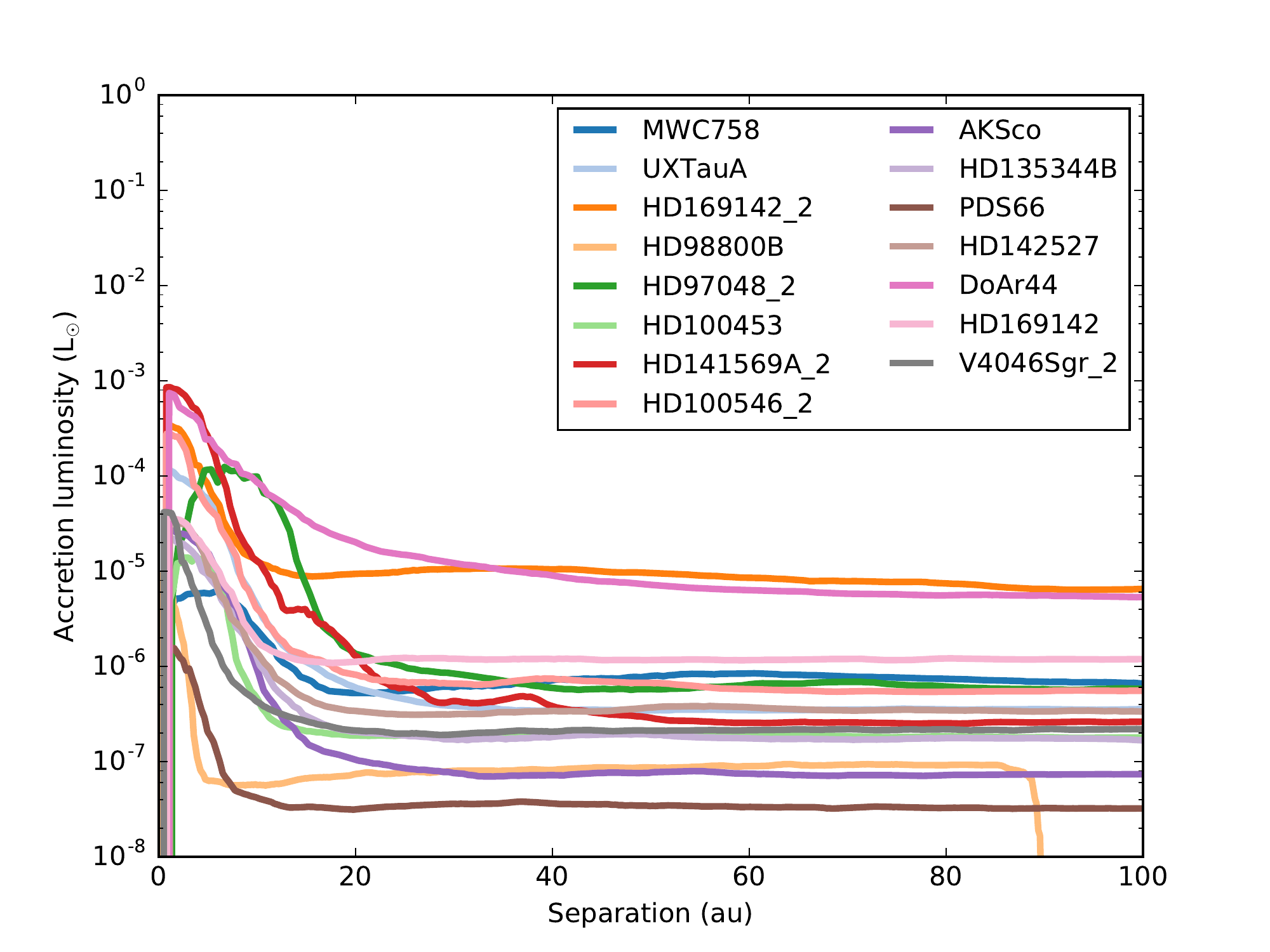}
\caption{Accretion luminosity limits for all the targets of the sample versus angular separation (left) and physical separation (right). }
\label{f:acc_l}
\end{center}
\end{figure*}

\begin{figure*}
\begin{center}
  \includegraphics[width=0.45\textwidth]{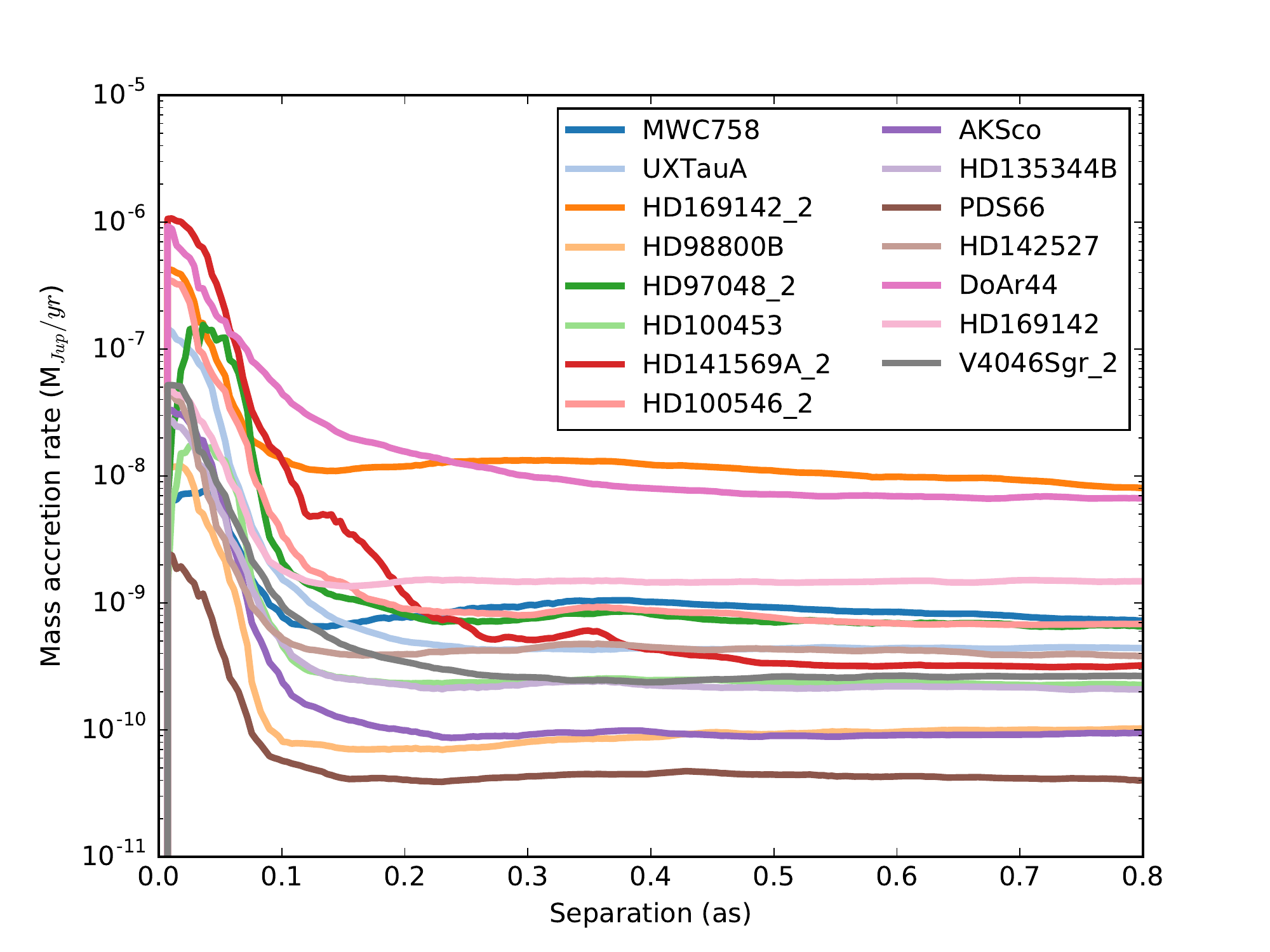}
  \includegraphics[width=0.45\textwidth]{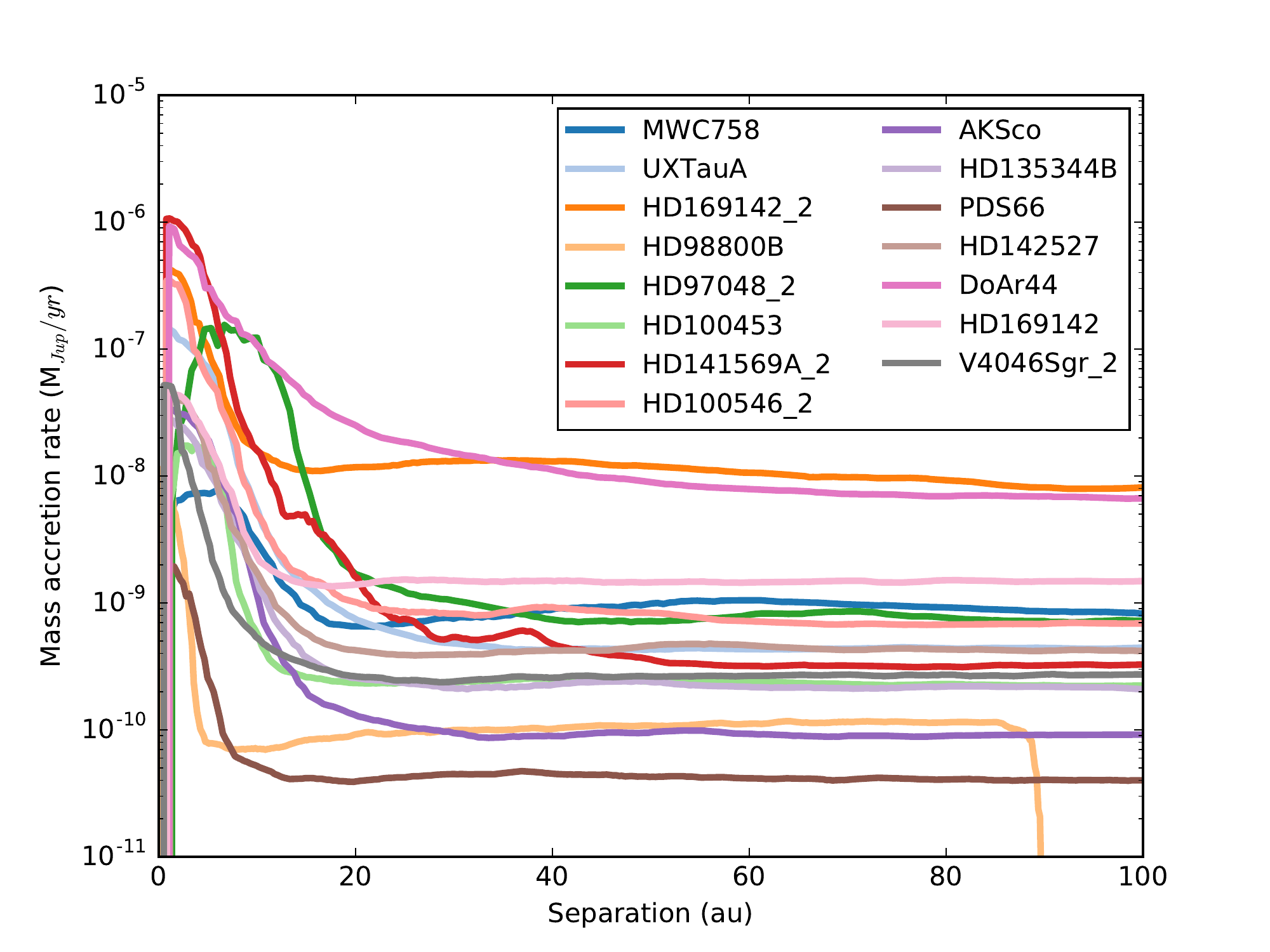}
\caption{{Mass accretion rate limits (for a 5 \MJup mass planet) for all the targets of the sample versus angular separation (left) and physical separation (right)}. }
\label{f:m}
\end{center}
\end{figure*}

Similarly to \citet{2019A&A...622A.156C}, in order to estimate the stellar flux in the continuum filter, we calculated the median of the count rate in an aperture of radius 1.5 arcsec and applied it in Eq. 4 from \citet{2017A&A...602A..53S} after subtracting the background calculated in a ring at r = 1\as \citep[see][]{2019A&A...631A..84M}. The filter zero point was also taken from \citet{2017A&A...602A..53S}. The calculated flux density was multiplied by the continuum filter effective width of 41.1 Angstrom \citep[see][]{2018A&A...619A...9S} in order to obtain the stellar flux. 
Subsequently, for each object, the contrast at each separation, which was calculated with respect to the PSF in the continuum images, was applied and the flux limit was directly calculated and converted into H$\alpha$ luminosity. The relationship between H$\alpha$ luminosity and accretion luminosity, as presented in \citet{2012A&A...548A..56R}, was used to obtain the upper limit for the accretion luminosity \citep[as in][]{2019A&A...622A.156C}. From simulations these formulas were also confirmed for planets \citep{2019ApJ...885...94T}.
In Fig.~\ref{f:flux}, we show the apparent flux limit for all the targets and the accretion luminosity in Fig.~\ref{f:acc_l}. In Fig.~\ref{f:m}, we show the mass accretion rate limits if we assume a planet of fixed mass 5 \MJup and 1.47 \RJup ({for the mean age of our sample}), and by exploiting the AMES-COND evolutionary models \citep{2001ApJ...556..357A} as in \citet{2019A&A...622A.156C}.{ The same Eq.~4 of \citet{2019A&A...622A.156C} was applied for the calculation. }

In summary: 
\begin{itemize}
\item For all the targets of our sample,  on average we reach H$\alpha$ line flux sensitivities, H$\alpha$ line luminosities and accretion luminosity upper limits of 2$\times$10$^{-15}$ erg/s/cm$^2$, 5$\times$10$^{-7}$ L$_{\odot}$, and 10$^{-6}$ L$_{\odot}$ respectively, beyond 0\farcs2.
  \item {We reach an average mass accretion rate sensitivity of 10$^{-9}$ \MJup/yr beyond 0\farcs2 for planets of 5 \MJup}.
\item The best sensitivities were obtained for AK\,Sco, HD\,142527, and PDS\,66 targets.
\item The worst sensitivity was obtained for DoAr\,44 because of the faintness of the central star (V=12.8), which is located at the limit of the adaptive optics (AO) system.
\item In half of the cases, we reached sufficient sensitivity to redetect PDS\,70\,b. 
\end{itemize}

\subsection{HD\,142527}
\label{s:HD142527}

HD\,142527 is a Herbig Ae/Be star that harbors a TD \citep[e.g.,][]{2013A&A...556A.123C,2017ApJ...840...60B,2017AJ....154...33A} and an M star accreting companion \citep{2012ApJ...753L..38B,2014ApJ...781L..30C,2018A&A...617A..37C, 2019A&A...622A..96C}. \citet{2018MNRAS.477.1270P} suggest that the shape of the disk has been carved by the interaction with the binary. \citet{2019A&A...622A..96C} retrieved the orbit for the companion using all the astrometric data available to date, they found a period of P = 35--137 yr; an inclination of i = 121--130 deg, and two families of values for the eccentricity: 0.2--0.45 and 0.45--0.7. In our data, the astrometry for the stellar companion is as follows: separation of 63.0 $\pm$ 1.5 mas and PA 96.4 $\pm$ 1 deg, which is consistent with the previous results from \citet{2019A&A...622A.156C} and \citet{2019A&A...622A..96C}.

\subsection{HD\,98800}
\label{s:HD98800}

HD\,98800 is a quadruple system at a distance of 47 pc \citep{2007A&A...474..653V}, which is composed of the following two binaries: HD\,98800\,BaBb, at a projected separation of 0\farcs8 from the spectroscopic binaries, and HD\,98800\,AaAb, which has a separation of $\sim$ 1 au \citep{2005ApJ...635..442B}. Around the B component, \citet{2007ApJ...664.1176F} found a transition disk composed of an optically thin component with an inner radius of 2 au, then a gap, and an optically thick component with a radius of 5.9 au. The presence of the gap might be explained by the carving of a planet. \citet{2019NatAs...3..278K} show ALMA results of the system, which present a very compact dust ring that is 2-au wide at a radius of 3.5 au. Because of a misunderstanding regarding the finding chart, the observations were carried out with the A component in the center of the image, and they will be performed again. {The two stellar companions (HD\,98800\,BaBb) of the spectroscopic binary were resolved in the visible for the first time.}

The two stellar components, which were resolved in the ZIMPOL images, have a separation with respect to the A star of 495.8 $\pm$ 1.5 mas and a position angle of 9.0 $\pm$ 1.0 deg, and a separation of 507.5 $\pm$ 1.5 mas, position angle of 7.5 $\pm$ 1.0 deg. They are separated by 17.7 mas. The brightest component (west component) has an H$\alpha$ flux of 6.5$\times$10$^{-16}$ L$_{\odot}$, an H$\alpha$ luminosity of 4.6$\times$10$^{-8}$ L$_{\odot}$, and an accretion luminosity of 1.1$\times$10$^{-8}$ L$_{\odot}$, while the faintest is not accreting.

 \section{Discussion and conclusions}
 \label{sec:conc}

 We conducted a survey of nearby (< 200 pc) TDs seen in the H$\alpha$ filter. The purpose was to look for accreting protoplanets in all the disks that present peculiar features that can be induced by the presence of companions. The survey, which was conducted with SPHERE/ZIMPOL, made use of the ASDI technique between the H$\alpha$ narrow filter and the adjacent continuum. Eleven targets were observed in excellent conditions (seeing < 0\farcs8) and the other three targets were added to the analysis for completeness. We did not detect any previously unknown companions, although the two components of HD\,99880\,BaBb were resolved for the first time in the visible.

 The contrast curves that we obtained for the targets of the survey reflect the exquisite performance of the instrument, and they give strong constraints on the presence of accreting companions around the selected stars.
If we compare our detection limits with the predictions from the population synthesis of \citet{2017A&A...608A..72M}, while also assuming complete cold accretion, we expect to be unable to spot low mass (< 1 \MJup) medium accreting ($\dot{M}$ = 10$^{-5 \pm 1}$ \MJup yr$^{-1}$) planets or high mass (1--15 \MJup) that are not heavily accreting ($\dot{M}$ < 10$^{-5.5}$ \MJup yr$^{-1}$).

The detection limits that we found can help us to investigate some properties of the putative protoplanets; the mass $M_p$, the planet radius $R_p$, and the planetary accretion rates $\dot{M}$. Assuming the accretion disk paradigm, it has been theorized that the $H\alpha$ luminosity, which is produced by shocks at some distance $R_s$ (magnetospheric radius), can be approximated by

\begin{equation}
L_{H\alpha} = 4 \pi R_s^2 \pi B_{\nu}(T) \nu_{H_{\alpha}} v_s/c,
\label{LH}
\end{equation}
where $\nu_{H_{\alpha}}$ is the $H\alpha$ line frequency, 
$v_s$ the free-fall velocity of the accreting gas, $c$ the speed of light, $B_{\nu}(T)$ the Planck function, and $T$ the temperature of the shock, which is generated by the gravitational conversion between gas accretion to accretion shock luminosity, that is, $L_{\rm shock} = \zeta G M_p  \dot{M} / R_p $, and where $\zeta = 1 - R_p / R_T$, where $R_T$ is the truncation radius \citep{Zhu2015}. { We used $R_T \sim 5R_P$ for the truncation radius, as in \cite{2019A&A...622A.156C}}.

{The shock temperature should be high enough to partially ionize the Hydrogen to be compatible with $H\alpha$ emission, therefore $T \sim 10^4 - 10^8$ K \citep{Aoyama2018}. The value of $v_s$ is crudely known, but it was estimated to be in the range $\sim 50 - 100 ~  \rm km ~ s^{-1}$ for accreting circumplanetary disks  \citep{Zhu2015, 2019ApJ...885L..29A}. Also, $v_s$ can be approximated by

\begin{equation}
v_s = (2 G M_p/R_p)^{1/2} \zeta^{1/2},
\label{Vs}
\end{equation}
where $R_p$ and $M_p$ are the radius and mass of the planet, respectively.}

Our detection limits give an upper boundary H$\alpha$ luminosity of $L_{H \alpha} = 5 \times 10^{-7} \rm L_{\odot}$.  
In assuming that the magnetospheric accretion shock occurs at the planet surface,
we constrained the maximum planetary-mass able to produce such emission by equating this upper-limit $L_{H\alpha}$ to Eq. \eqref{LH}.
{For instance, if we assume a shock temperature of $T = 10^4 \rm ~ K$ 
and $v_s = 50, ~ 100 ~  \rm km ~ s^{-1}$, we obtain from Eq. \eqref{LH} (assuming $R_T = 5 R_p$); $R_p = 1.3,~ 0.9 ~  \RJup$, respectively.  Using the resulting $R_p$, we derived the following planetary masses from Eq. \eqref{Vs}; $1.1,~ 3.3 ~ \MJup$, respectively, making the second example with high free-fall velocity less realistic. }


It has yet to be determined if some of the nondetections are due to the intrinsic variability of protoplanets, which can be highly variable over time, changing their luminosity on short timescales on the order of hours. That could be a possible explanation for the nondetection of these kinds of objects, despite the very high sensitivity that we can reach with the current instrumentation. If this is the case, in the future, multiepoch opbservations are necessary to unveil the nature of accreting protoplanets in transition disks.

Another possible explanation for the nondetections could be that accreting planets are located within $\sim$200 mas from the host star, where our observation are not sensitive enough due to the bright central speckles. Also, one has to consider that the assumptions that we are making in order to derive accretion luminosities from line luminosities may be inappropriate. For example, the estimate of the PDS\,70\,b accretion rate, using the different accretion luminosity estimate by \citet{2019ApJ...885L..29A}, was much higher than the one presented in \citet{2018ApJ...863L...8W} for the same line luminosity. This could mean that a given accretion rate would correspond to a lower line luminosity if more realistic conversions are applied.

\begin{acknowledgements}
{We thank the anonymous referee for the constructive comments that improved the manuscript.}  A.Z. acknowledges support from the CONICYT + PAI/ Convocatoria nacional subvenci\'on a la instalaci\'on en la academia, convocatoria 2017 + Folio PAI77170087. G.C. thanks the Swiss National Foundation for the financial support under the grant number 200021\_169131. M.M. acknowledges financial support from the Chinese Academy of Sciences (CAS) through a CAS-CONICYT Postdoctoral Fellowship administered by the CAS South America Center for Astronomy (CASSACA) in Santiago, Chile, and support from Iniciativa Cient\'ifica Milenio via the N\'ucleo Milenio de Formaci\'on Planetaria. 
N.H. has been partially funded by the Spanish State Research Agency (AEI) Project No. ESP2017-87676-C5-1-R and No. MDM-2017-0737 Unidad de Excelencia {\em Mar\'{\i}a de Maeztu} - Centro de Astrobiolog\'{\i}a (INTA-CSIC).
\end{acknowledgements}

 
\bibliographystyle{aa}
\bibliography{halpha}

\end{document}